\documentclass[journal]{IEEEtran}

\usepackage{graphicx}
\usepackage{epsfig}
\usepackage{dcolumn}
\usepackage{bm}
\usepackage{amsmath}
\usepackage{amssymb}
\usepackage{upquote}
\usepackage{tabularx}
\usepackage{textcomp}

\usepackage{listings}
\usepackage{widetext}

\begin{document}

\title{Reliable SPICE Simulations of Memristors, Memcapacitors and Meminductors}

\author{Dalibor~Biolek, {\it Member, IEEE}, Massimiliano~Di~Ventra and Yuriy~V.~Pershin, {\it Senior Member, IEEE}
\thanks{D. Biolek is with the Department of Electrical Engineering/Microelectronics, University of Defence/Brno University of Technology, Brno, Czech Republic \newline
e-mail: dalibor.biolek@unob.cz}
\thanks{M. Di Ventra is with the Department of Physics, University of California, San Diego, La Jolla, California 92093-0319 USA \newline e-mail: diventra@physics.ucsd.edu.}% %<-this % stops a space
\thanks{Y. V. Pershin is with the Department of Physics and Astronomy and USC Nanocenter, University of South Carolina, Columbia, SC, 29208 USA \newline e-mail: pershin@physics.sc.edu.}}% <-this % stops a space}

%\thanks{Manuscript received October XX, 2011; revised March YY, 2012.}


%\markboth{Journal of \LaTeX\ Class Files,~Vol.~6, No.~1, January~2007}%
%{Shell \MakeLowercase{\textit{et al.}}: Bare Demo of IEEEtran.cls for Journals}

\maketitle

\begin{abstract}
\boldmath Memory circuit elements, namely memristive, memcapacitive and meminductive systems, are gaining considerable
attention due to their ubiquity and use in diverse areas of science and technology. Their modeling within the
most widely used environment, SPICE, is thus critical to make substantial progress in the design and
analysis of complex circuits. Here, we present a collection of models of different memory circuit elements
and provide a methodology for their
accurate and reliable modeling in the SPICE environment. We also provide codes of these models written in the most popular SPICE versions (PSpice, LTspice, HSPICE) for the benefit of the reader.
We expect this to be of great value to the growing community of
scientists interested in the wide range of applications of memory circuit elements.
\end{abstract}

%\begin{IEEEkeywords}
%Memristors, Memcapacitors, Meminductors, Analog circuits, Emulators
%\end{IEEEkeywords}

% \fi
\IEEEpeerreviewmaketitle

\section{Introduction}
There is presently a large interest in what are commonly called memristors, memcapacitors and meminductors (or
collectively simply {\it memelements}), namely resistors, capacitors and inductors with memory, respectively~\cite{diventra09a}. This class of circuit elements offers considerable advantages compared to traditional devices. Specifically, these are {\it two-terminal} electronic devices that can store {\it analog} information even in the absence of a power source. From the point of view of potential applications, memelements open up the possibility of manipulating and
storing information within a totally different computing paradigm \cite{diventra13a,Alibart10a,pershin11d,pershin12a,linn2012beyond,thomas2013memristor},
 extend functionality of traditional devices \cite{Sacchetto12a},
 as well as serve as model systems for certain biological processes and systems \cite{pershin09b,erokhin2010bio,Johnsen11a,traversa13a}.

Mathematically, an $n$th-order $u$-controlled memelement is
defined by the equations~\cite{diventra09a}
\begin{eqnarray}
y(t)&=&g\left(x,u,t \right)u(t) \label{Geq1}\\ \dot{x}&=&f\left(
x,u,t\right) \label{Geq2}.
\end{eqnarray}
Here, $u(t)$ and $y(t)$ are any two circuit variables (current,
charge, voltage, or flux) denoting input and output of the system,
$x$ is an $n$-dimensional vector of internal state variables, $g$ is
a generalized response, and $f$ is a continuous $n$-dimensional
vector function. Special interest is devoted to devices determined
by three pairs of circuit variables: current-voltage (memristive systems),
charge-voltage (memcapacitive systems), and flux-current (meminductive systems).
Two other pairs (charge-current and voltage-flux) are linked
through equations of electrodynamics and therefore are of no
practical interest. Devices defined by the relation of charge and
flux (the latter being the integral of the voltage) are not considered as
a separate group since such devices can be redefined in the
current-voltage basis~\cite{chua71a}.

However, future progress in the analysis of complex circuits involving any of these elements requires reliable simulation tools that are easy to implement and flexible enough to provide solid predictions on a wide range of physically realizable models. The Simulation Program with Integrated Circuit Emphasis (SPICE) environment is one such general-purpose simulator that has been successfully used in the analysis of integrated circuits for forty years. SPICE allows the testing of complex circuits before they are actually implemented experimentally, thus saving a lot of time and resources in their fabrication.

Being new on the circuit scene, memelements do not have many years of testing within the SPICE environment. Nonetheless, more and more SPICE models are being considered with different levels of complexity \cite{Biolek2009-1,Biolek2009-2,Benderli2009-1,Rak10a,Biolek10b,Biolek11b,Kolka11a,pershin12c,Kvatinsky13a}.
Oftentimes, readers are interested in the SPICE code itself and its reliability within the range
of physical parameters used. Unfortunately, both the codes and reliability criteria are not always
available in the literature thus limiting the use of some of the most popular SPICE models of memelements.

This paper attempts to fill this gap by providing several models of ideal and non-ideal memristive,
memcapacitive and meminductive elements and their implementation (codes) in the most popular SPICE versions (PSpice, LTspice, HSPICE), focusing on the well-known PSpice. Our goal is also to provide a general methodology for accurate modeling within this environment so that readers interested in implementing different models can easily build from the examples we provide in this paper and venture out on their own.
We think this could also serve as an excellent teaching tool complementing others (e.g., experiment-based ones \cite{Pershin12b}) for the next generation of scientists and engineers interested in this field. This methodology is given in Section \ref{methodology} which follows this Introduction. In later Sections we will then focus on specific examples of memristive, memcapacitive and meminductive systems and their modeling in SPICE.

Importantly, instead of focusing on different levels of sophistication in describing the {\it same}
electronic device, we concentrate on SPICE models of physically {\it different} memory devices (e.g., bipolar, unipolar, etc.) that are generally classified as
memristive, memcapacitive or meminductive systems.
For completeness, such a presentation
is integrated with models of ideal memory elements -- memristors, memcapacitors and meminductors.
For each device, we select a reasonable complexity in modeling essential features of device operation relying, in some cases,
on original models proved to be useful in device simulations.

\section{Methodology for accurate and reliable modeling of memelements with SPICE} \label{methodology}

Throughout the development of memelement models and their implementation in SPICE-family simulation programs, several limitations and specific features of these programs should be taken into consideration. This way situations can be avoided in which the program finds a solution which is burdened with errors, either evident or not apparent at first sight, or when the solution is not found at all. The above two kind of problems, i.e., imperfections and non-convergence issues, can be magnified in circuits containing memelements, i.e., which have specific hysteresis behavior. For example, it is shown in \cite{kolka2013frequency} that the classical algorithms of finding the periodical steady states, which are implemented in several simulation programs such as HSPICE RF, Micro-Cap, and partially in LTspice, can be ineffective for circuits containing memelements. In addition, the work \cite{Tetzlaff13a} calls attention to the fact that the periodic solution of the circuit containing the classical model of the HP memristor \cite{Biolek2009-1,Biolek2009-2,Benderli2009-1,Rak10a,strukov08a,joglekar09a}, found within the transient analysis, can be entirely corrupted via common numerical errors accumulated throughout the analysis. Nevertheless, without an extended analysis, these results can be easily accepted as correct.

Paradoxically, problems with precision and reliability can also arise when working with the ideal memelement models whose behavior is free from the ubiquitous parasitic effects. Such simplification can produce poor conditions for the operation of SPICE computational core. On the other hand, the analysis of the behavior of such ideal models is of great importance, if understanding the fundamental properties of memelements is the key aim of the simulation. Clearly, any deviation from the ideal behavior due to parasitic effects is undesirable and troublesome.

The SPICE modeling and simulation is about the compromise between accuracy of the results and the speed and reliability of the procedure to obtain them. Since the accuracy of the analysis of memelements is frequently a key factor, it is advisable to build the model just in relation to this criterion. If convergence problems appear, such model should be modified, taking into account the well-known rules of the reliable behavioral modeling \cite{kundert1995designer}, combining them with proper settings of the program options and the parameters of concrete analysis \cite{kielkowski1998inside}.

The transient analysis is the most widely used SPICE analysis of circuits containing memelements. That is why we focus on the rules on how to build such memelement models in SPICE which would comply with specific limitations of the numerical algorithms used throughout the transient analysis in the SPICE environment. Some of these rules should be applied with the aim of achieving results as accurate as possible. The purpose of other rules is to prevent  convergence problems while analyzing the circuits with memelements, or to solve them as early as they appear.

The mathematical model of each memelement can be divided into the submodel of the element port (of memristive, memcapacitive, or meminductive nature), and into the part modeling the differential equations for the internal state variables which control the port parameters (the memristance, memcapacitance, and meminductance). Both groups are modeled in SPICE environment via a mix of the tools of conventional and behavioral modeling. The behavioral modeling uses especially the controlled sources and mathematical formulae. The accuracy and reliability of the simulation results depend on the following factors which are then discussed below:

\begin{description}
  \item[$\bullet$] Numerical limits, given by a finite precision and finite dynamic range of the number representation in SPICE environment.
  \item[$\bullet$] Rules of building-up behavioral models, resulting in continuous equations and their derivatives, bearing in mind the numerical limits.
  \item[$\bullet$] The way of modeling the state and port equations.
  \item[$\bullet$] Setting the parameters of transient analysis and the global parameters.
\end{description}

The recommendations discussed below are applicable to a wide class of SPICE-family simulation programs. Some specifics of concrete programs are analyzed separately. Details which are beyond this text can be found in the program documentation, e.g., \cite{HSPICE08a,PSpice11a,Micro10a}.

\subsection{Numerical limits affecting accuracy and convergence in SPICE-family programs}

Double-precision binary floating-point (a ``double'' in short) is a commonly used format on PCs, enabling the number representation within the dynamic range from $2^{-1022}$ to $2^{1023}$, thus from about $10^{-308}$ to $10^{308}$. The significant precision is $53$ bits with $52$ explicitly stored, which gives about 16 digits of accuracy. The maximum relative rounding error (the machine epsilon) is $2^{-53}$, i.e., approximately $10^{-16}$. In SPICE environment, this format shares all voltages and currents and also the system variable TIME used throughout the transient analysis.
However, the above limits are modified by concrete SPICE-family programs. For example, PSpice limits the voltages and currents larger than $10^{10}$ Volts and Amps and the maximum derivatives are $10^{14}$. These limits are rather higher in HSPICE, LTspice and Micro-Cap. The smallest nonzero numbers which the programs can process are not commonly documented. For example, it is $10^{-30}$ for PSpice.
The above limits together with other items, which are defined in global settings (acceptable relative and absolute errors, number of iterations, etc.) affect the accuracy but also the program (in)ability to find the solution within these limits.

\subsection{Rules of building-up behavioral models} \label{sec2B}
Some of the rules are well documented in the literature \cite{kundert1995designer,kielkowski1998inside,HSPICE08a,PSpice11a,Micro10a}. Below is given a brief account with reference to the memelement modeling for the subsequent transient analysis. Specific details are omitted. They appear in Section \ref{secC}.

\subsubsection{Components with (un)realistic parameters} \label{secB1}

Behavioral modeling of non-electric quantities in SPICE, based on various analogies, for example modeling of the position of the boundary between the doped and undoped layers of a TiO$_2$ memristor, can lead to the selection of atypical values of the parameters of the elements in the substitutive electric circuit. As a result, the computed voltages and currents can be extremely high or low, causing numerical difficulties.
It is useful to avoid small floating resistors because any error in the computed nodal voltages of such resistors results in large error currents \cite{kundert1995designer}. If the resistor was included in the circuit as a current probe, then it should be replaced by a 0-Volt voltage source. Note that a large number of such probes increases the size of the circuit matrix which can negatively influence the program operation. Similar difficulties as small floating resistors can arise with large floating capacitors.
Also note that convergence problems can appear in the feedback systems with large loop gains. Some modeling techniques use passive R, C, and L elements with negative parameters. These methods are not recommended because they can cause unstable behavior of the model.

\subsubsection{(Dis)continuous models}

Discontinuous models result from the operation of several memory elements, for example memristive systems with threshold \cite{pershin12c} or multi-state memristor switching memories with discontinuous memristance versus state characteristics \cite{chua2011resistance}. The rigorous modeling of these discontinuities is thus desirable for providing high precision of the model. On the other hand, it is a potential source of numerical problems which can cut down the precision. A possible strategy, which can work well especially for not so large-scale systems, is to model rigorously the discontinuous characteristics of memory elements in the first step. In the case of convergence problems or unrealistic results, some of the techniques of smoothing the characteristics can be applied subsequently. For example, the step function (STP in PSpice, U in LTspice), which is frequently used for modeling the saturations inside memdevices, can be replaced by a sigmoid function with adjustable parameters, which sets the maximum possible slope of the transition between two states. The IF function for modeling piece-wise constitutive relations of memelements, can be modeled such that the derivatives are not changed abruptly in order to remove the discontinuities of the first derivatives at the corner points.
The signal waveforms can serve as other sources of discontinuities. The well-known conventions should be followed here, for example that the pulses should be modeled with realistic rise/fall times.

\subsubsection{Models (in)sensitive to numerical errors}

Models of some analog circuits are highly sensitive to numerical errors which originate from a finite precision of the number representation, and which can be due to specific operations of computational algorithms. The model, built up from such blocks, can then behave differently in the environments of various simulation programs, even if the simulations run under apparently identical conditions. The simulation outputs can be far away from the real behavior of the systems being modeled. However, it is entirely up to the user to notice it. The errors are obvious in several cases but not always.

It is also necessary to distinguish the source of the model sensitivity: it can be either the nature of the modeled circuit or the improper way of constructing the mathematical model.
The models with extremely long time constants exhibit high sensitivities to numerical errors, which work as accumulators of these errors during the transient simulation run where the differential equations are solved numerically. A typical example of a sensitive circuit is an ideal integrator which is, however, the basic building block of ideal memristors, memcapacitors, and meminductors. Any numerical problem at arbitrary instants of time during the integration algorithm of the transient analysis run can then influence the results computed at all the subsequent instants.
A more important source of numerical problems can be the block of time-domain differentiation. It does not work as an accumulator but as an amplifier of the truncation errors, with unlimited bandwidth since its gain increases by 6 dB with doubling the frequency.

The $\textnormal{d}/\textnormal{d}t$ operation should be avoided in behavioral modeling, for example via a substitution of the $\textnormal{d}/\textnormal{d}t$-type model by its dual integrating version (see Section B.6). As an interesting consequence, the capacitor currents and inductor voltages are not computed in SPICE as accurately as the capacitor voltages and inductor currents. For example, the capacitor current is proportional to the differentiation of voltage with respect to time. Then any numerical error in the voltage is amplified to the current waveform. This suggests a useful rule: as far as possible, we should prefer computations within the behavioral models with capacitor voltages and inductor currents rather than with capacitor currents or inductor voltages.

Since the above circuits either accumulate or amplify errors, the only thing we can do against such effects is to minimize the consequences, for example via selecting a proper integration method and tuning its parameters (see Section \ref{secC}).
On the other hand, the model sensitivity to numerical errors can be undesirably increased via an improper construction of the model. For example, if the model gain is spread unreasonably among individual cascade blocks, it can bring the local attenuation of the signal near the low limit of the dynamic range of the number representation or, on the contrary, its overflow. Another typical case is an improper subtraction of two commensurate numbers which results in a high truncation error. An example of this is the well-known Joglekar window function for modeling nonlinear dopant drift in TiO$_2$ memristors, which for the parameter $p = 1$ \cite{joglekar09a} can be written in two following ways:
\begin{equation}
f\left( x\right) = 1-\left( 2x-1 \right)^2
\end{equation}
or
\begin{equation}
f\left(x\right) = 4x\left( 1-x\right).
\end{equation}

For the memristor in its boundary state with a maximum memristance, when $x$ is close to $0$, the first model generates significantly larger errors. Due to the finite dynamic range of the double format, the term $(2x-1)^2$ cannot differ from 1 by less than the value of $2^{-53}$. Then one can conclude that for all values $x<2.776\times 10^{-17}$ the values of window function are cut to zero. For the second model, however, such limitation appears if $x$ is less than its minimum value for the double type, i.e., for $x<2^{-1022}=2.225 \times 10^{-308}$.
Such a model sensitivity to truncation errors can play a detrimental role within all commonly used models of  memelements which utilize window functions (see Section \ref{SecB4}).

\subsubsection{Selection of state variables of memelements - the key to accurate computation} \label{SecB4}

Truncation errors and their accumulation throughout the integration process of the transient analysis can be the cause of mistaken results even for the simulation of simple circuits containing memelements. The reason can be in an improper form of the differential state equation(s) of the memelement which results in high sensitivity of its solution to the truncation errors. It is shown in \cite{Tetzlaff13a} that such high sensitivity occurs for the well-known differential equation of the TiO$_2$ memristor where the time-domain derivative of the normalized position $x$ of the boundary between the doped and undoped layers is directly proportional to the memristor current and the window function $f(x)$, which tends to zero at boundary points $x=0$ and $x=1$. If the memelement approaches very closely the boundary state, then SPICE can erroneously evaluate, due to the truncation errors, that this state is already attained. Then the memelement state is frozen since the derivative of the state variable with respect to time is zero. The element can change from this state only due to some other numerical errors. In doing so, however, the duration of this ``pseudo-fixed'' state, which is of a random character, can significantly affect subsequent computations.

The fact that something is wrong with the simulation results is obvious only when it is found that some memelements fingerprints are violated. This is of particular concern because it takes effect latently and without any warnings or error messages of the simulation program. However, it can corrupt the simulation results for complex circuits with other memelements utilizing the window functions, such as memcapacitors \cite{Biolek10b} and meminductors \cite{Biolek11b}. For cases when the element state is swept far from the boundary states, the simulation is correct. However, it fails when trying to simulate, for example, the hard switching effects.

The above troubles can be avoided via a selection of a more suitable state variable which would lead to another differential equation. Its solution must be much less sensitive to numerical errors. Evaluating this state variable, the memelement parameter, for instance the memristance, is computed in the second step, either directly from the state variable-to-parameter relationship, or by the medium of the state variable which has caused troubles in the classical approach. It is shown in \cite{Tetzlaff13a} that the so-called native state variable (for example the charge or flux for the memristor), is the good choice for modeling ideal memelements. Then the state equation is a simple model of ideal integrator. It is a potential accumulator of the truncation errors though, but the resulting effect is much better than for the above sensitive case.

\subsubsection{Behavioral modeling of integrators}
The model of the integrator is necessary for modeling the state equations. SPICE implementation of the integrator is usually in the form of a grounded 1-Farad capacitor with a controlled current source in parallel. If the source current is equal to the quantity which is integrated, then the capacitor voltage in Volts is equal to the computed integral during the transient analysis. The initial state at time 0 can be set via the IC attribute of the capacitor. Shunt resistor with a large resistance, not disturbing the integration process, is necessary for providing DC path to the ground.

Note that extremely high capacitances can generate non-convergence issues. The integration capacitance can be decreased simultaneously with decreasing charging current. Then it is useful to analyze if this current, which models the quantity being integrated, has realistic values. Otherwise, the numerical problems at the bottom area of the dynamic range can take effect.

Several SPICE-family programs offer built-in functions for signal integration, for example the SDT function in PSpice and Micro-Cap and the IDT function in LTspice. The properties of these functions are not documented. It is proved for PSpice Cadence v. 16.3 that the SDT function accumulates the truncation errors slightly more than the conventional integrator model. In other words, both models provide the same accuracy if a smaller step ceiling is used for the integration via SDT function.
The precision of the integration process also depends on the parameters of transient analysis, on the integration method, and on other simulator options (see Section \ref{secC}).

\subsubsection{Modeling memristive, memcapacitive, and meminductive ports}

These ports are modeled as R, C, and L two-terminal devices with varying parameters. For example, the memristor is modeled as a resistor whose resistance is controlled by the state quantity. The model of more general memristive systems can use a resistor with nonlinear current-voltage characteristic which is controlled by a set of state variables. Similar structures can be used for modeling memcapacitive and meminductive systems, utilizing capacitors and inductors with varying characteristics.
The SPICE standard does not support a direct modeling of $R$, $C$, and $L$ elements with varying parameters. Apart from  specific features of several programs, these elements can be modeled indirectly via tools of behavioral modeling, namely with the help of the controlled sources and mathematical formulae.

\vspace{0.3cm}

\noindent {\bf Memristive systems}

Resistors with varying resistance $R$ or conductance $G$ are modeled either as voltage source controlled by the equation $V=R(x,I,t)I$, where $I$ is the source current, or as a current source controlled by the formula $I=G(x,V,t)V$, where $V$ is the source voltage and $x$ are internal state variables. Several rules should be followed:

1) During the simulation, the source formulae should not generate any divisions by small numbers, let alone zero, and they should not generate other numerical errors (for example, any subtraction of commensurate numbers which is sensitive to rounding errors). If the memristance of the modeled device is close to zero, it is more preferable to work with the memristance than with the memductance, and to use the model based on the voltage, not the current source.

2) If it is possible to divide the formula for the modeled memristance or memductance into fixed and variable parts, then the fixed part can be modeled by a classical fixed element and the remaining part by a behavioral controlled source. The variable part should comply with the above rule 1). The fixed part must represent positive value of the memristance or memductance. This provides reliable models of the memristive/memconductive port via Th\'{e}venin/Norton models without any potential conflicts due to such connections of ideal sources violating the Kirchoff's voltage law/ Kirchoff's current law.

Note that several SPICE-family programs enable a direct modeling of resistors via equations. In HSPICE, the resistance can be a function of arbitrary voltage or current, or of any other system variable such as TIME. Similar features are provided also by Micro-Cap.

\vspace{0.3cm}

\noindent {\bf Memcapacitive systems}

The capacitive port of charge-controlled memcapacitive systems can be modeled by the formula $V=D(x,q,t)q$, where $q$ is charge and $D$ is inverse of the memcapacitance, which depends on the state variables $x$ and on the charge. This implies that such port can be modeled via a voltage source with the voltage computed from the state variables and the charge. The charge is calculated as the integral of the port current.

Accordingly, the capacitive port of voltage-controlled memcapacitive systems can be modeled as $q=C(x,V,t)V$, where $C$ is a memcapacitance, which depends on state variables $x$ and on the voltage. It appears from this that such port can be modeled via a controlled charge source. Nevertheless, such a source is not commonly available in all the SPICE-family programs. Then the current should be computed via differentiating the charge with respect to time, and the capacitive port should be implemented by the current source. However, the differentiation is not suggested as a reliable numerical procedure.

It is advisable to follow the rules No. 1) and 2) for the memristance modeling, with the appropriate modifications for the memcapacitive model. In the case of partitioning the (inverse) memcapacitance into the fixed and varying parts, the capacitive port can be modeled by a fixed capacitor and controlled source in (series) parallel.

Note that some SPICE-family programs enable more general modeling of the capacitors. Micro-Cap provides the capacitance definition via a formula, or the capacitor charge can be described as a function of the capacitor voltage. LTspice can model the capacitor charge as a general function of a special variable $x$ which is the capacitor voltage.  HSPICE enables the capacitance definition as a function of its terminal voltages, external voltages and currents, or their combinations (HSPICE RF), or the capacitor charge can be defined as a function of the terminal and other voltages and currents. Also, some present versions of OrCAD/Cadence PSpice  can work with the charge sources, namely through the extended syntax of the G-type controlled source, which uses a formula for the charge. Such programs enable convenient modeling of memcapacitive systems, controlled via the current or voltage.

The following rule should be applied when working with the memcapacitive models: every node must have its DC path to ground. If it is not the case, a large shunting resistor must be added to the circuit such that its resistance cannot affect the simulation.

\vspace{0.3cm}

\noindent {\bf Meminductive systems}

The inductive port of voltage- (or flux)-controlled meminductive systems can be modeled by the formula $I=\Lambda(x,\phi,t)\phi$, where $\phi$ is flux linkage and $\Lambda$ is the inverse of meminductance, which depends on the state variables $x$ and on the flux. This implies that such port can be modeled via a controlled current source. The current can be calculated from the state variables and the flux, the latter one via integrating of the port voltage.

Accordingly, the inductive port of current-controlled meminductive systems can be modeled as $\phi=L(x,I,t)I$, where $L$ is the meminductance which depends on the state variables $x$ and on the current. Such port can be modeled via a controlled flux source. Nevertheless, such a source is not commonly available in all the SPICE-family programs. Then the voltage should be computed via differentiating the flux with respect to time, and the inductive port should be implemented by the voltage source. Remember that the differentiation is not a preferred procedure.
In the case of partitioning the (inverse) inductance into the fixed and varying parts, the inductive port can be modeled by a fixed inductor and controlled source in (parallel) series.

Several SPICE-family programs enable more general modeling of the inductors, thus they can be recommended for a more comfortable modeling of current-controlled meminductive systems. Micro-Cap provides the inductance definition via a formula. Alternatively, the inductor can be defined by a flux formula which must depend on the inductor current. LTspice can model the inductor flux as a general function of a special variable $x$, which is the inductor current. HSPICE enables the inductance definition as a function of nodal voltages and branch currents. The inductor can be also defined by the flux formula. Present OrCAD/Cadence PSpice versions use special F-syntax of the E-type controlled source (the flux source), which generates the voltage as a time-derivative of the flux. The flux can be defined by a formula.

If the convergence or other numerical problems appear due to the inductors in the circuit, the rule should be applied that all inductors should have a parallel resistor, which limits the impedance at high frequencies. The resistance must be high enough in order to prevent its influence to the circuit parameters. Its value should be set equal to the inductor's impedance at the frequency at which its quality factor begins to roll off. The purpose of such resistor is to prevent undesirable voltage spikes associated with abrupt changes of the inductor current, causing the convergence problems.
Also note that the SPICE programs do not allow the loops containing only ideal voltage sources and inductors. Such loops must be completed by resistors. Corresponding resistances must be low enough but not extremely low (see Section \ref{secB1}).

\section{SPICE modeling of memristive devices} \label{sec3}

\subsection{Model {\normalfont R.1:} Ideal memristor} \label{secR1}

{\bf Model:} In a current-controlled memristor \cite{chua76a}, the memristance $R$ depends only on charge, namely,
\begin{equation}
V_\text{M}=R(q(t))I \label{r1:eq1}
\end{equation}
with the charge related to the current via time derivative $I=\textnormal{d}q/\textnormal{d}t$. The direct use of
Eq. (\ref{r1:eq1}), however, is uncommon. More common are models inspired by physics of resistance switching. In particular, a popular model \cite{strukov08a} is based on the assumption that the memristive device consists of two regions (of a low and high resistance) with a moving boundary. The total memristance can be written as a sum of resistances of two regions
\begin{equation}
R(x)=R_{\text{on}}x+R_{\text{off}}(1-x). \label{r1:eq2}
\end{equation}
Here, $x\in [0,1]$ parameterizes the position of boundary, and $R_{\text{on}}$ and $R_{\text{off}}$ are limiting values of memristance. The equation of motion for $x$ can be written, for example, using a window function $W(x)$ as
\begin{equation}
\frac{\textnormal{d}x}{\textnormal{d}t}=k W  (x) I, \label{r1:eq3}
\end{equation}
where $k$ is a constant, and $W(x)$ is often selected as \cite{joglekar09a}
\begin{equation}
W(x)=1-\left( 2x-1 \right)^{2p}, \label{r1:eq4}
\end{equation}
where $p$ is a positive integer number.

\begin{figure}[tb]
 \begin{center}
   \includegraphics[width=0.45\textwidth]{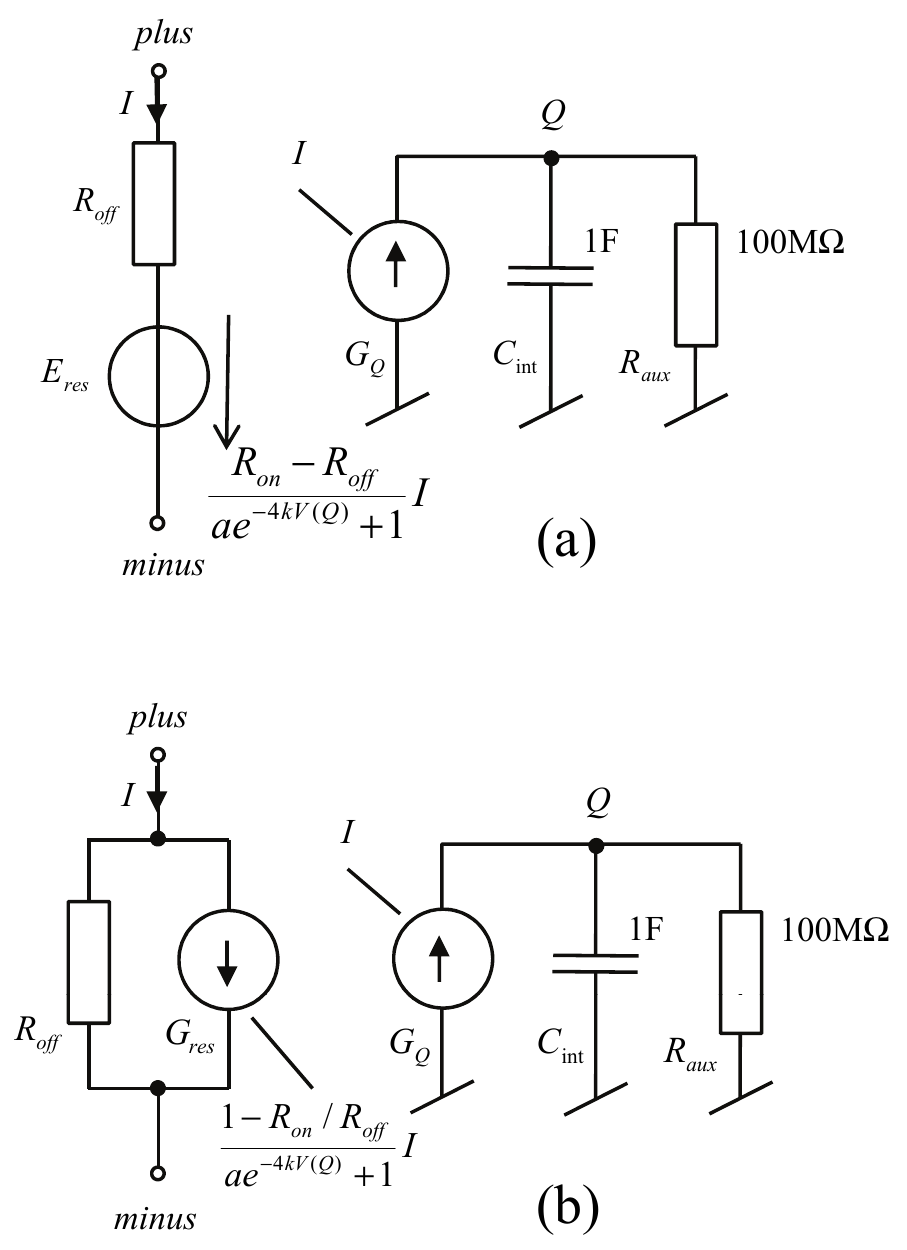}
\caption{\label{figR1a} Possible SPICE implementations of the ideal memristor model (Eqs. (\ref{r1:eq1}), (\ref{r1:eq7})). The memristive port can be modeled via a voltage source with a serial resistor (a) or via an equivalent current source with a parallel resistor (b). Here, $V(Q)$ is the voltage of the node $Q$, which has the same numerical value in Volts as the charge $q(t)$ in Coulombs.}
\end{center}
\end{figure}

{\bf Features:} The above model takes into account boundary values of memristance. It does not involve a switching threshold, is not stable against fluctuations, and exhibits over-delayed switching \cite{diventra13b}. We emphasize that Eqs. (\ref{r1:eq2})-(\ref{r1:eq4}) describe an ideal current-controlled memristor. In principle, Eq. (\ref{r1:eq3}) can be integrated for an arbitrary function $W(x)$ and thus $x$ can be expressed as a function of $q$. For example, if $W(x)$ is given by Eq. (\ref{r1:eq4}) with $p=1$, then one finds
\begin{equation}
\frac{1}{4}\textnormal{ln}\frac{x}{1-x}=k(q(t)+q_0) \label{r1:eq5}
\end{equation}
where $q_0$ is the integration constant (initial condition). Consequently,
\begin{equation}
R(q(t))=R_\text{off}+\frac{R_\text{on}-R_\text{off}}{e^{-4k(q(t)+q_0)}+1}. \label{r1:eq6}
\end{equation}
It can be more convenient to re-write $q_0$ in terms of the initial memristance $R_{ini}= R(q=0)$ resulting in
\begin{equation}
R(q(t))=R_\text{off}+\frac{R_\text{on}-R_\text{off}}{ae^{-4kq(t)}+1},\;\;\; a=\frac{R_\text{ini}-R_\text{on}}{R_\text{off}-R_\text{ini}} . \label{r1:eq7}
\end{equation}
Equation (\ref{r1:eq7}) represents a reliable model for SPICE simulation: the memristance is derived as a function of the native state variable $q$, thus the state equation is not sensitive to the truncation errors in contrast to Eq. (\ref{r1:eq3}). In SPICE, the charge can be obtained via integrating the port current $I$ by the capacitor $C_{int}$ according to Fig. \ref{figR1a}. Then the charge in Coulombs is equal to the voltage of the node $Q$ in Volts. It is obvious from Eq. (\ref{r1:eq7}) that the memristive port can be modeled as a serial connection of the fixed $R_\text{off}$ resistor and a controlled voltage source (see Fig. \ref{figR1a} (a)). For modeling large circuits, which can be prone to convergence problems, the Norton equivalent according to Fig. \ref{figR1a} (b) can be more advantageous. For the sake of brevity, only the codes of the first model are given in the Appendix \ref{app_R1}.

\begin{figure}[tb]
 \begin{center}
    \includegraphics[width=9.0cm]{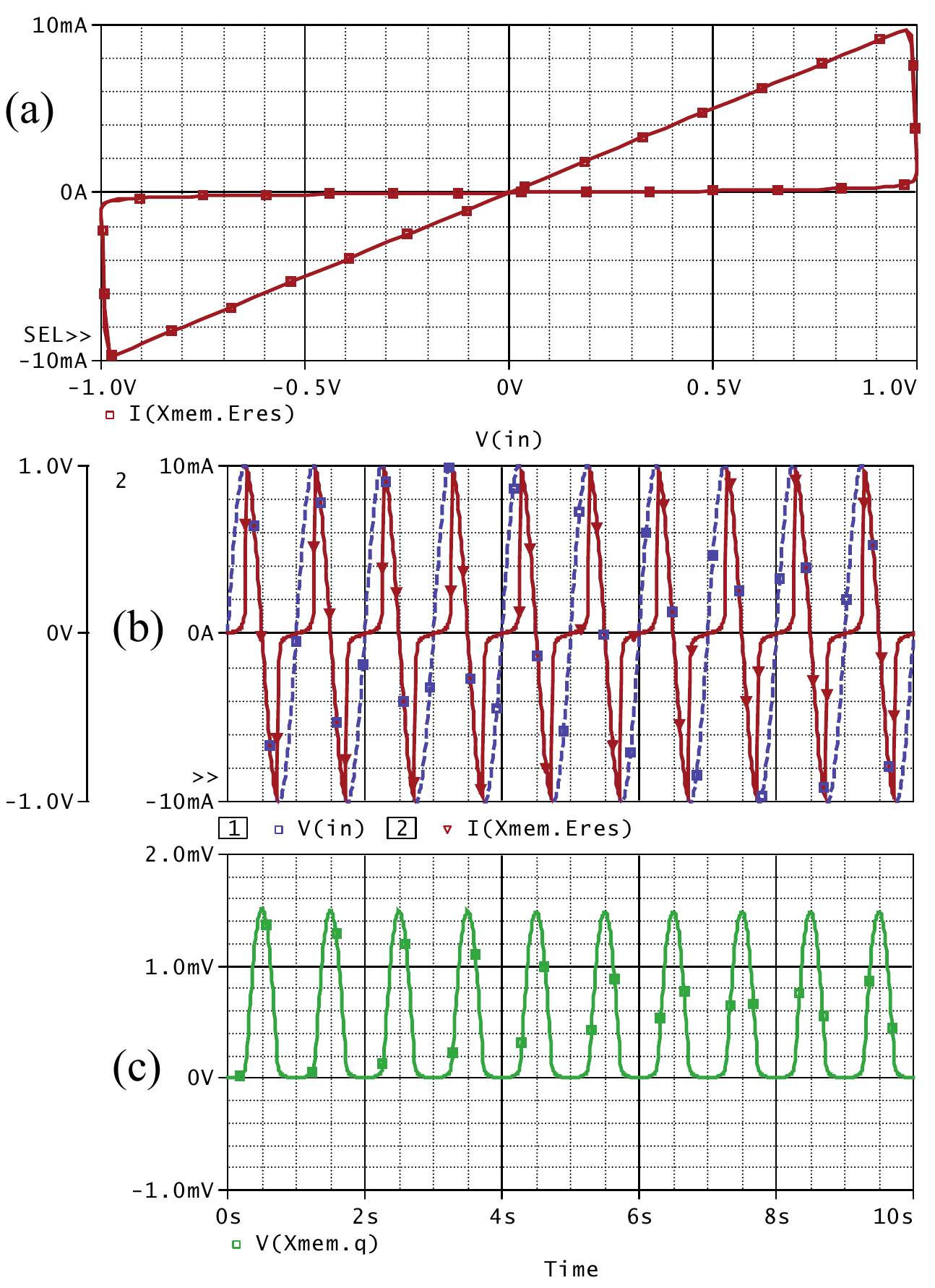}
\caption{\label{figR1b} PSpice outputs for the case of an ideal memristor driven by the sine-wave 1V/1Hz voltage source: (a) current-voltage pinched hysteresis loop, (b) voltage and current waveforms, and (c) charge (i.e., integral of current) waveform.}
\end{center}
\end{figure}

{\bf Results:} Figure \ref{figR1b} shows the simulation results in PSpice for the memristor model from the Appendix \ref{app_R1}, utilizing the circuit file therein. The correctness of the results can be evaluated via the charge waveform (i.e., the voltage of the internal node $Q$ of the subcircuit) which must be periodical without any initial transients. For LTspice, it is preferable to use Gear integration which leads to the best results. Note that PSpice user cannot select the Gear method.

We emphasize that the memristive port can be modeled in HSPICE also by a direct formula:

\begin{lstlisting}
Rmem plus minus R=
+'Roff+(Ron-Roff)/(1+a*exp(-4*k*V(q)))'
\end{lstlisting}

However, the accuracy of the computation cuts down. It can be increased back by decreasing the maximum time step.

\subsection{Model {\normalfont R.2:} Bipolar memristive system with threshold} \label{secR2}

\begin{figure}[tb]
 \begin{center}
    \includegraphics[width=5cm]{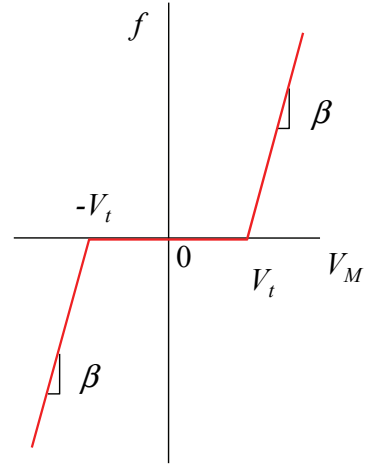}
\caption{\label{figR2a} Sketch of the function $f(V_\text{M})$ modeling the voltage threshold property.}
\end{center}
\end{figure}

{\bf Model:} Several approaches to take into account a threshold-type switching are available in the literature \cite{pershin09b,pickett2009switching,yakopcic2011memristor,Kvatinsky13a}. Here, we consider a model of a voltage-controlled memristive system with voltage threshold suggested in Ref. \cite{pershin09b} by two of us (YVP and MD). For the sake of simplicity, we consider its reduced version (without switching below the threshold) \cite{pershin13a}. In this model, the memristance $R$ plays the role of the internal state variable $x$, namely, $x \equiv R$, defining the device state via the following equations
\begin{eqnarray}
I&=&x^{-1}V_\text{M}, \label{r2:eq1} \\
\frac{\textnormal{d}x}{\textnormal{d}t}&=&f(V_\text{M})W(x,V_\text{M}) \label{r2:eq1a}
\end{eqnarray}
where $f(.)$ is a function modeling the device threshold property (see Fig. \ref{figR2a}) and $W(.)$ is a window function:
\begin{eqnarray}
f(V_\text{M})&=&\beta \left( V_\text{M}-0.5\left[ |V_\text{M}+V_\text{t}|-|V_\text{M}-V_\text{t}| \right]\right) ,\label{r2:eq1b} \\
W(x,V_\text{M})&=&\theta\left( V_\text{M}\right) \theta\left(
R_\text{off}-x\right)+ \theta\left(- V_\text{M}\right) \theta\left(
x-R_\text{on}\right) . \;\;\;\;\; \label{r2:eq2}
\end{eqnarray}
Here $\theta(\cdot)$ is the step function, $\beta$ is a positive constant
characterizing the rate of memristance change when $|V_\text{M}|> V_\text{t}$,
$V_\text{t}$ is the threshold voltage, and  $R_\text{on}$ and $R_\text{off}$ are limiting
values of the memristance $R$. In Eq. (\ref{r2:eq2}), the role of $\theta$-functions
is to confine the memristance change to the interval between
$R_\text{on}$ and $R_\text{off}$.

\begin{figure}[tb]
 \begin{center}
    \includegraphics[width=8cm]{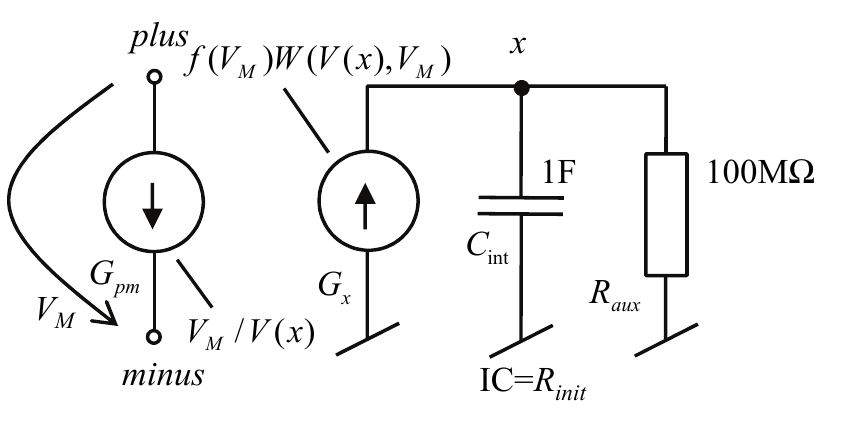}
\caption{\label{figR2b} SPICE model of the memristive device with threshold.}
\end{center}
\end{figure}

{\bf Features:} Eqs. (\ref{r2:eq1})-(\ref{r2:eq2}) provide a compact realistic description of bipolar memristive devices. The model takes into account boundary values of memristance and threshold-type switching behavior.  In many real memristive devices, the resistance change is related to the atomic migration induced by the applied field and not by the electric current flow. Therefore, models with voltage threshold \cite{pershin09b,yakopcic2011memristor} are physically better justified than those with the current one \cite{pickett2009switching,Kvatinsky13a}. From the point of view of the numerical analysis of Eq. (\ref{r2:eq1}), the division by the state variable $x$ is not a problem since the memristance varies only within the $R_{\text{on}}$ to $R_{\text{off}}$ limits. Based on Eqs. (\ref{r2:eq1}) and (\ref{r2:eq1a}), the basic schematics of SPICE implementation is presented in Fig. \ref{figR2b}.

In this approach, the derivative of the memristance (\ref{r2:eq1a}) is modeled by the current of the controlled source $G_x$, and its integral –- the memristance in Ohms –- is equal to the voltage of the node $x$ in Volts. According to Eq. (\ref{r2:eq1}), the memristive port is modeled by the current source $G_{pm}$. Its current is computed as a ratio of the terminal voltage and the memristance.
Equations (\ref{r2:eq1b}) and (\ref{r2:eq2}) contain discontinuous function (step) and function with discontinuous derivatives (absolute value). It can be a source of serious convergence problems, especially for applications utilizing large-scale models. In such cases, smoothed functions can be used based on sigmoid modeling of the step function according to the formula
\begin{equation}
\theta_\text{S}(x)=\frac{1}{1+e^{-x/b}}
\end{equation}
where $b$ is a smoothing parameter.

Then the smoothed version of the absolute value function, $\text{abs}_\text{S}(x)$, can be
\begin{equation}
\text{abs}_\text{S}(x)=x\left[\theta_\text{S}(x)-\theta_\text{S}(-x) \right].
\end{equation}
If a convergence problem appears, a proper tradeoff between the accuracy and reliability can be usually found via tweaking the $b$ parameter. For the simplicity, the corresponding smoothed functions $\text{stp}_\text{S}(x)$, $\text{abs}_\text{S}(x)$ and the functions $f_\text{S}(x)$ and $W_\text{S}(x)$ derived from them are defined in the source codes \ref{app_R2} directly within the individual subcircuits.

\begin{figure}[tb]
 \begin{center}
    \includegraphics[width=9cm]{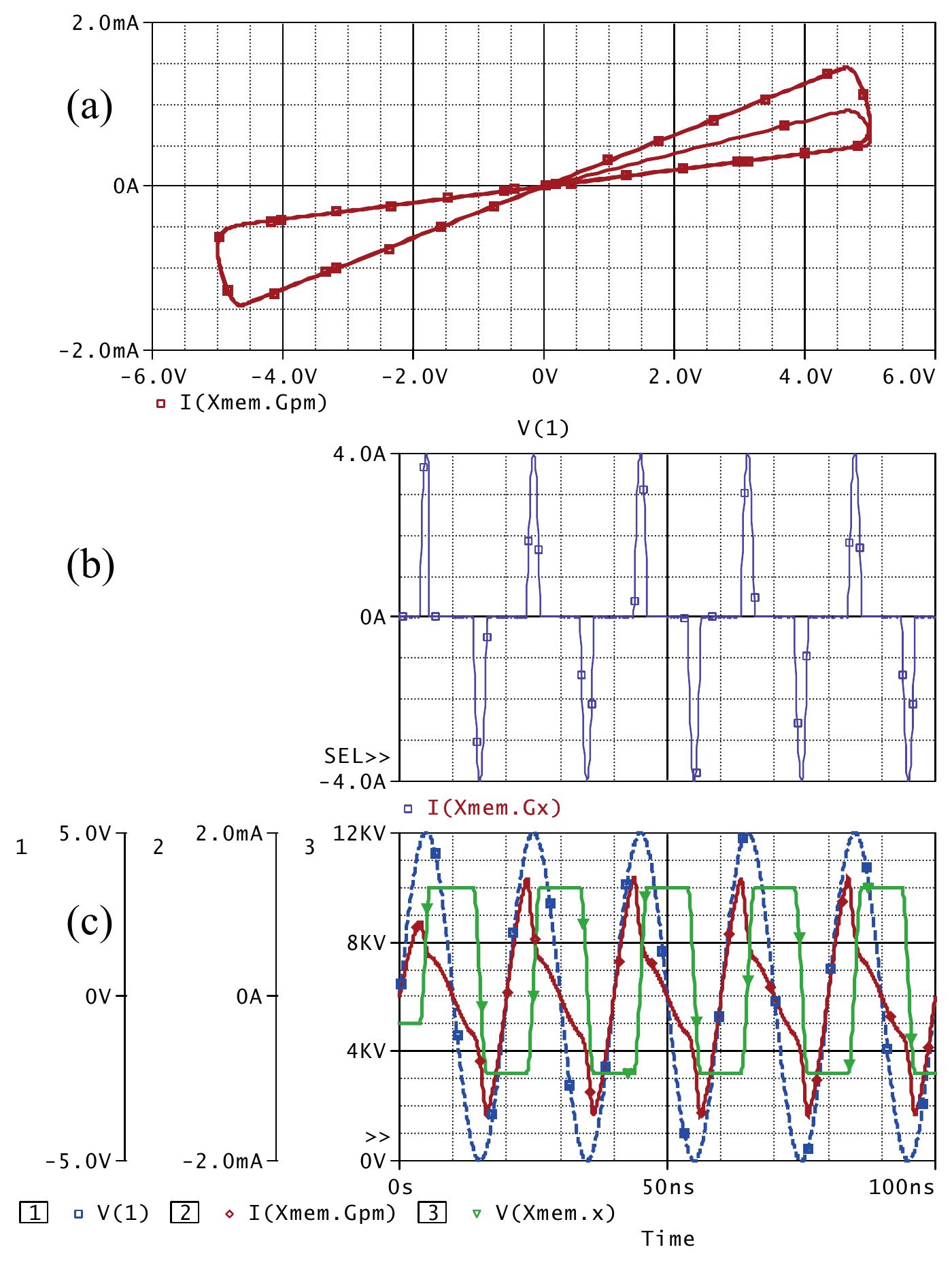}
\caption{\label{figR2c} PSpice outputs for the memristive device with threshold driven by a sine-wave excitation. The parameters are defined in the SPICE code in the Appendix \ref{app_R2}.}
\end{center}
\end{figure}

{\bf Results:} Examples of the PSpice outputs, generated from the source codes from the Appendix \ref{app_R2}, are shown in Fig. \ref{figR2c}. As follows from Fig. \ref{figR2a}, the function $f(V_\text{M})$ generates narrow pulses when the memristive device is excited by sine-wave voltage $V_\text{M}$ with the amplitude $V_\text{max}>V_\text{t}$. Considering the positive pulse in Fig. \ref{figR2c}, it will be integrated into the voltage of the node $x$ until the memristance $R=V(x)$ approaches its boundary value $R_{\text{off}}$. At this instant, the window function $W$ and also the current of the source $G_x$ are set to zero, and the memristance is fixed to the value $R_{\text{off}}$. This state persists until the voltage $V_\text{M}$ drops below the negative threshold level $-V_\text{t}$. Then the function $f(V_\text{M})$ becomes negative. It causes the negative current pulse of the source $G_x$, and its integral will decrease the memristance towards $R_{\text{on}}$. It is obvious from Fig. \ref{figR2c} that, although the memristance did not drop to its bottom limit, the current is cut off at the instant when the voltage $V_\text{M}$ has exceeded the threshold $–V_\text{t}$ (the effect of the window $W$). The memristance is held on the low level all the time when the voltage $V_\text{M}$ travels within the stable zone between both threshold levels. Then the system continues in the motion in the frame of its periodical steady state.

It follows from the above analysis that the combination of unreasonably time step and error criteria can result in an incorrect determination of the boundary conditions in the integration of current pulses. If this happens, then the simulated waveforms can be distorted due to significant errors. One can make certain of this via step-by-step selection of various parameters/options of the transient analysis or error criteria. To identify incorrect results or to achieve the necessary accuracy, we can use the following guides (they are true for the specific netlist in the Appendix \ref{app_R2}):
\begin{enumerate}
\item{The upper level of the memristance (the curve V(Xmem.x)) must be $R_{\text{off}}$. Each declination from this value is a numerical error.}
\item{The bottom value of the memristance (if it does not reach the boundary $R_{\text{on}}$, see Fig. \ref{figR2c}), must be}
\end{enumerate}
\begin{equation}
R_{\text{off}}-\frac{\beta}{2\pi f}V_\text{t}\left[2\sqrt{\left(\frac{V_{max}}{V_\text{t}}\right)^2-1}-\pi+2\sin^{-1}\frac{V_\text{t}}{V_{max}} \right] \label{eq16}
\end{equation}
where $f$ is the signal frequency.

For the simulation example from Fig. \ref{figR2c}, the necessary accuracy can be accomplished e.g. via a low relative error RELTOL=1u in combination with the maximum time step equal to one thousandth of the simulation time. Then for PSpice results in Fig. \ref{figR2c}, the low-level memristance is 3.1819 k$\Omega$ whereas the accurate value according to (\ref{eq16}) is 3.1847 k$\Omega$. Note that the simulation in HSPICE according to code \ref{app_R2} provides even more accurate computation. If the simulation program enables to select the integration method, then the Gear integration is preferable in this case due to its stability throughout the analysis over many repeating periods.

Note that the current of the source $G_x$ in the SPICE code \ref{app_R2} is multiplied by a number 1p and that the integrating capacitor has the capacitance of 1 pF. It is due to the optimization of the dynamic range of the source current. Without this multiplication, the current would reach extreme values of 4 TA, which is not optimal with regard to the standard analysis options. In addition, since the voltage of the node $x$ in volts is equal to the memristance in Ohms, this voltage appears in kiloVolts, being also out of the typical values. That is why, if necessary,  the following optimization step would lead to set and compute the memristance in kiloOhms, not in Ohms, with an increase of the capacitance $C_{int}$ by three orders to 1nF. Then the voltage $V(x)$ would appear on the common level of Volts. HSPICE provides the most accurate results among all three simulation programs. The option RUNLVL=6 forces HSPICE into the regime of enhanced precision (see Section \ref{secC}).

\subsection{Model {\normalfont R.3:} Phase change memristive system} \label{secR3}

{\bf Model:} In phase change memory (PCM) cells \cite{burr2010phase}, the information storage is based on the reversible
phase transformation of relevant materials. In terms of memristive formalism, PCM cells can be described
as unipolar second-order current- or voltage- controlled memristive systems. Following general ideas of Ref. \cite{dao2011compact}, we consider here a simple model of PCM cells
based on equations describing thermal and phase change processes. Using the temperature $T$ and the crystalline fraction $C_\text{x}$ as internal state variables, the model of PCM cells can be written as
\begin{eqnarray}
I&=&R^{-1}(C_\text{x},V_\text{M})V_\text{M}, \label{r3:eq1} \\
\frac{\textnormal{d}T}{\textnormal{d}t}&=&\frac{V_\text{M}^2}{C_\text{h}R(C_x,V_\text{M})}+\frac{\delta}{C_\text{h}} \left(
T_\text{r}-T\right), \label{r3:eq2} \\
\frac{\text{d}C_\text{x}}{\text{d}t}&=&\alpha \left(1-C_\text{x} \right) \theta\left(T-T_\text{x}\right)\theta\left(T_\text{m}-T\right)\nonumber \\
&&-\beta C_\text{x}\theta\left(T-T_\text{m}\right), \label{r3:eq3}
\end{eqnarray}
where
\begin{equation}
R(C_\text{x},V)=R_\text{on}+\left( 1-C_\text{x} \right)\frac{R_\text{off}-R_\text{on}}{e^\frac{V-V_\text{t}}{V_0}+1}, \label{r3:eq4}
\end{equation}
$C_\text{h}$ is the heat capacitance, $\delta$ is the dissipation
constant, $T_\text{r}$ is the ambient temperature,
$\theta [.]$ is the step function, $T_\text{m}$ is the melting point,
$T_\text{x}$ is the glass transition point, $\alpha$ and $\beta$ are constant defining
 crystallization and amorphization rates, respectively, $V_\text{t}$ is the threshold voltage, $R_\text{on}$ and $R_\text{off}$ are limiting values of memristance, and
 $V_0$ is parameter determining the shape of $I-V$ curve.

{\bf Features:} This simple model takes into account crystallization (when $T>T_\text{m}$) and amorphization (when $T_\text{m}>T>T_\text{x}$) processes neglecting, however, a negative differential resistance region close to the threshold voltage $V_\text{t}$. Although this region can be straightforwardly incorporated into the model (written in the current-controlled form), it is not important for the memory cell operation (reading/writing voltages are always beyond that region).
Several other approaches to model PCM cells in SPICE are available \cite{cobley2006parameterized,ventrice2007phase,sonoda2008compact}.

\begin{figure}[tb]
 \begin{center}
    \includegraphics[width=9cm]{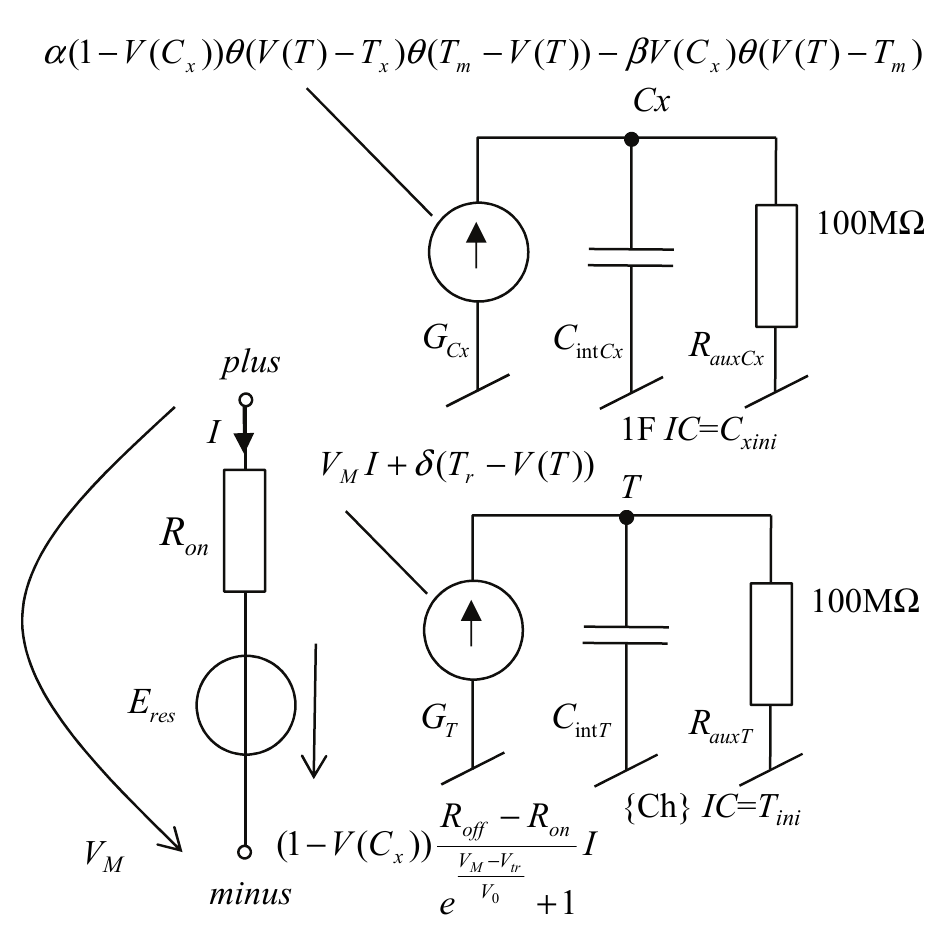}
\caption{\label{figR3a} Model of phase change memristive system for SPICE implementation.}
\end{center}
\end{figure}

The schematic in Fig. \ref{figR3a} represents three submodels of the phase change memory: the submodel of the resistive port ($R_{\text{on}}$, $E_{res}$) according to Eqs. (\ref{r3:eq1}) and (\ref{r3:eq4}), and submodels of integrators for computing the temperature ($G_T$, $C_{intT}$, $R_{auxT}$) and $C_x$ ($G_{Cx}$, $C_{intCx}$, $R_{auxCx}$) according to Eqs. (\ref{r3:eq2}) and (\ref{r3:eq3}). Note that the power $V_\text{M}^2/R$  in Eq. (\ref{r3:eq2}) dissipated on the memristive port can be also computed as a product of voltage and current as shown in Fig. \ref{figR3a}.

\begin{figure}[tb]
 \begin{center}
    \includegraphics[width=9cm]{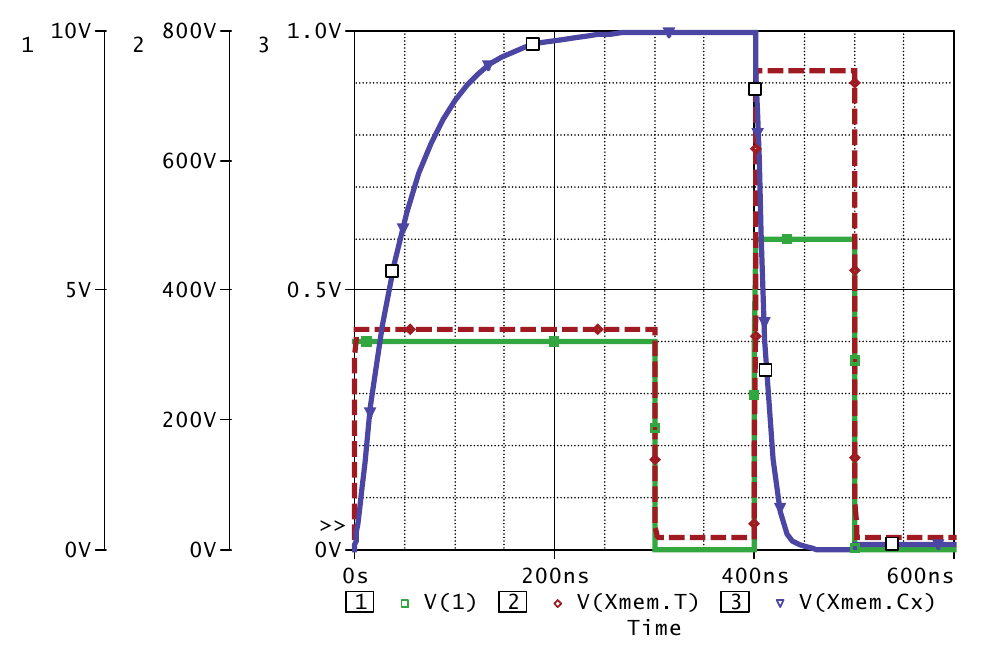}
\caption{\label{figR3b} PSpice results of transient analysis of PCM excited by voltage pulses V(1).}
\end{center}
\end{figure}

{\bf Results:} The transient analysis results provided by PSpice code from Appendix \ref{app_R3} are given in Fig. \ref{figR3b}. The 4V/300ns voltage pulse sets the temperature to ca 339$^\circ$C, i.e. above the crystallization temperature, which causes the transition to the crystalline phase (see the transition in $C_x$ from 0 to 1). The second 6V/100ns pulse sets the temperature to ca 739$^\circ$C, which is above the melting point, and the crystalline fraction $C_x$ drops close to zero.

\subsection{Model {\normalfont R.4:} Insulator-to-metal transition memristive system} \label{secR4}

{\bf Model:} A model \cite{pickett2012sub} of insulator-to-metal phase transition
device employs the metallic phase fraction expressed in radial coordinates, $u=r_\text{met}/r_\text{ch}$, as an internal
state variable. The model equations are
\begin{eqnarray}
V&=&R_\text{ch}(u)I \label{r4:eq1} \\
\frac{\textnormal{d}u}{\textnormal{d}t}&=&\left( \frac{\textnormal{d}\Delta H}{\textnormal{d}u} \right)^{-1}\left( R_\text{ch}(u)I^2-\Gamma_\text{th}(u)\Delta T\right),  \label{r4:eq2}
\end{eqnarray}
where
\begin{eqnarray}
R_\text{ch}(u)&=&\frac{\rho_\text{ins}L}{\pi r^2_\text{ch}}\left[1+\left(\frac{\rho_\text{ins}}{\rho_\text{met}} -1\right)u^2 \right]^{-1},  \label{r4:eq3} \\
\Gamma_\text{th}(u)&=&2\pi L \kappa \left( \textnormal{ln}\frac{1}{u}\right)^{-1},  \label{r4:eq4} \\
\frac{\textnormal{d}\Delta H}{\textnormal{d}u}&=&\pi L r_\text{ch}^2\left[\hat{c}_p\Delta T \frac{1-u^2+2u^2\textnormal{ln}u}{2u(\textnormal{ln}u)^2}+2\Delta\hat{h}_\text{tr}u \right]. \label{r4:eq5} \;\;\;\;\;\;
\end{eqnarray}
Here, $r_\text{met}$ is the radius of metallic core, $r_\text{ch}$ is the conduction channel radius, $H$ is the enthalpy, $\Gamma_\text{th}$ is the thermal conductance of the insulating shell, $\rho_\text{ins}$ is the insulating phase electrical resistivity, $\rho_\text{met}$ is the metallic phase electrical resistivity, $L$ is the conduction channel length, $\kappa$ is the thermal conductivity, $\hat{c}_p$ is the volumetric heat capacity, $\Delta\hat{h}_\text{tr}$ is the volumetric enthalpy of transformation. Typical values of model parameters can be found in Ref. \cite{pickett2012sub}.

{\bf Features:} This model describes unipolar current-controlled memristive device based on a thermally-driven insulator-to-metal phase transition. As demonstrated in \cite{pickett2012sub}, it provides realistic modeling of complex dynamic behavior of the device including sub-nanosecond switching times. On the other hand, the structure of Eqs. (\ref{r4:eq4}) and (\ref{r4:eq5}), containing logarithms of the phase composition state variable $u$, divisions by $u$, and divisions by logarithm of $u$, where $u$ can vary between 0 and 1, indicates potential numerical problems. To prevent them, it is useful to provide artificial limitations of the variable $u$ in SPICE code.

\begin{figure}[tb]
 \begin{center}
    \includegraphics[width=9cm]{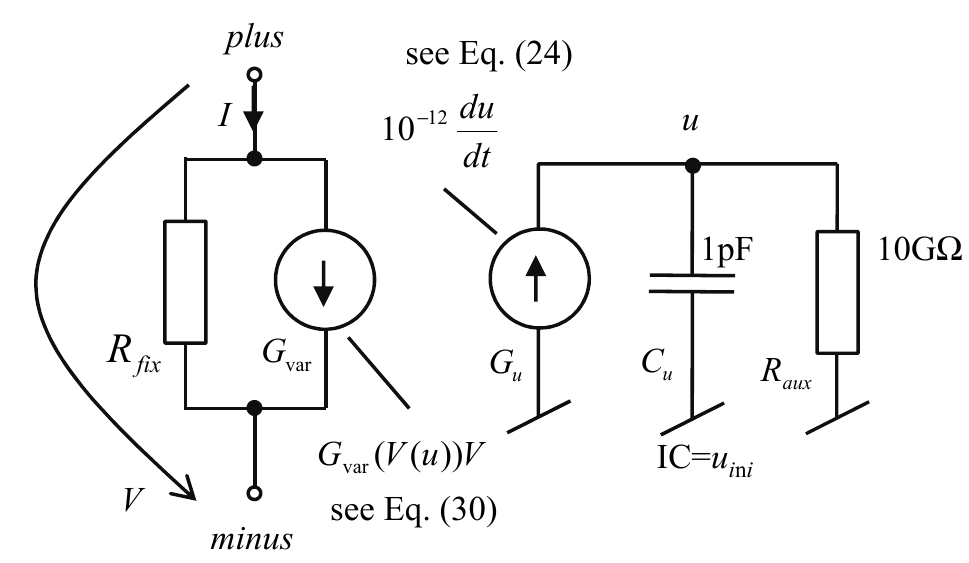}
\caption{\label{figR4a} Model of phase change memristive system for SPICE implementation.}
\end{center}
\end{figure}

Equations (\ref{r4:eq1}) and (\ref{r4:eq3}) can be rewritten in the form
\begin{eqnarray}
I&=& R_\text{fix}^{-1}V+G_\text{var}(u)V, \\
R_\text{fix}&=&\frac{\rho_\text{ins}L}{\pi r_\text{ch}^2}, \\
G_\text{var}(u)&=&\frac{\pi r_\text{ch}^2}{L}\left( \frac{1}{\rho_\text{met}}-\frac{1}{\rho_\text{ins}}\right) u^2.
\end{eqnarray}
The corresponding modeling of the memristive port via a parallel combination of a resistor $R_{fix}$ and a controlled current source $G_\text{var}$ is shown in Fig. \ref{figR4a}. The variable $u$ is found through the integration of the right-side of Eq. (\ref{r4:eq2}) using a capacitor $C_u$ which is charged by a current source $G_u$. Since the time derivatives of $u$ come up to high values, a proper scaling by the factor of $10^{-12}$ is provided according to Fig. \ref{figR4a} to prevent convergence problems.

\begin{figure}[tb]
 \begin{center}
    \includegraphics[width=9cm]{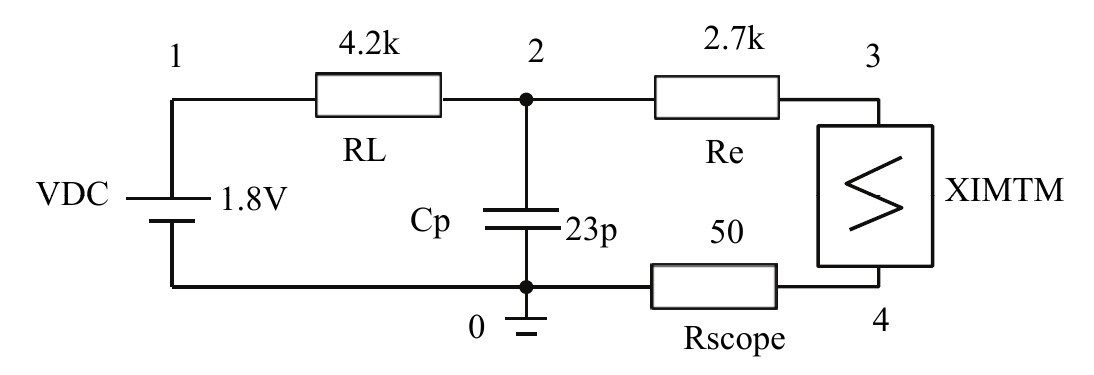}
\caption{\label{figR4b} Insulator-to-metal transition memristive system (XIMTM) as a part of the relaxation oscillator \cite{pickett2012sub}.}
\end{center}
\end{figure}

\begin{figure}[tb]
 \begin{center}
    \includegraphics[width=9cm]{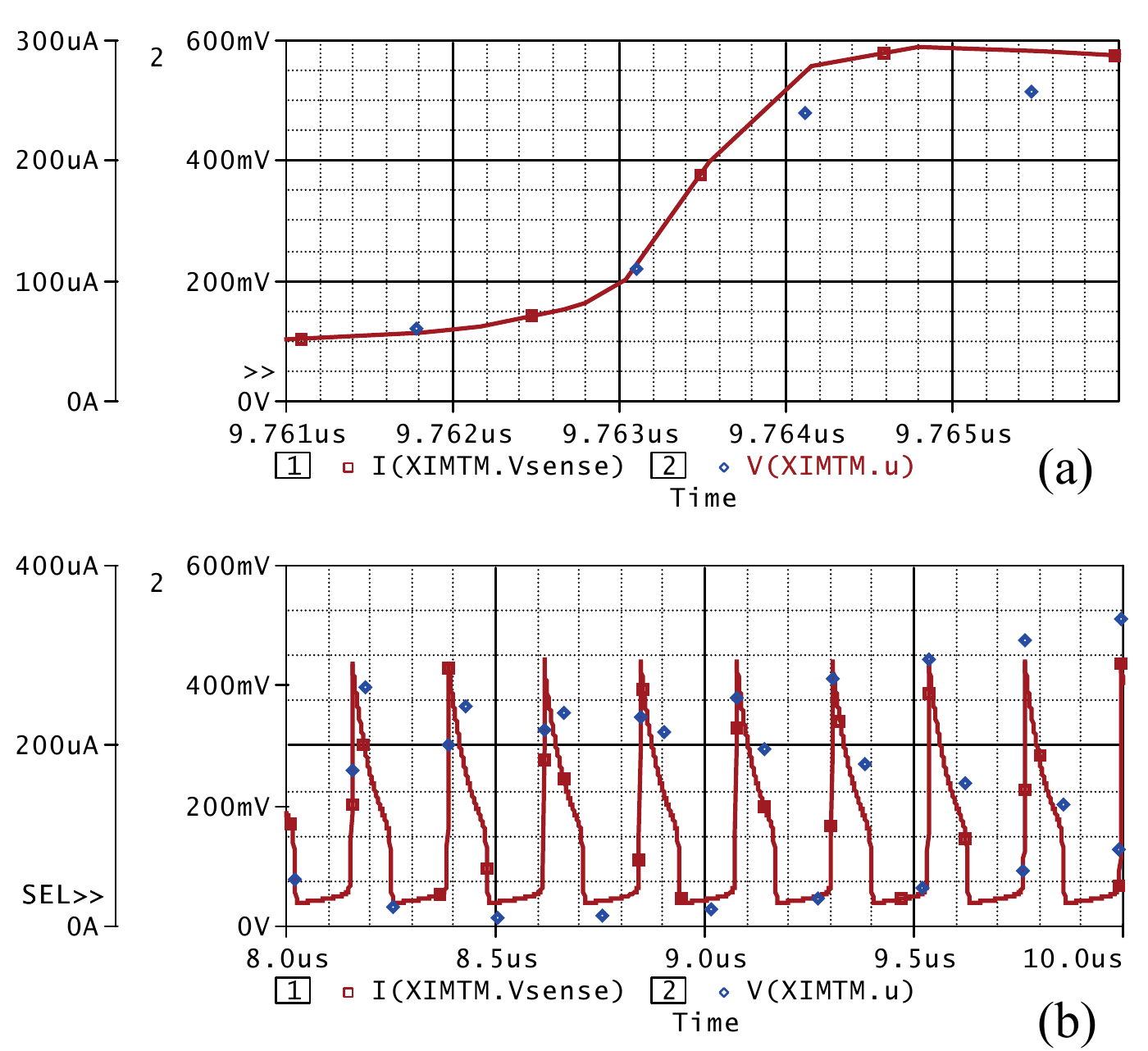}
\caption{\label{figR4c} Transient analysis of circuit from Fig. \ref{figR4b} in PSpice: current pulses through the memristive system (solid lines), phase composition state variable $u$ (dashed lines).}
\end{center}
\end{figure}

{\bf Results:}  For demonstrating the features of the corresponding SPICE model R.4 in Appendix \ref{app_R4}, the simulation of the experimental Pearson-Anson relaxation oscillator, described in \cite{pickett2012sub}, has been performed. As shown in Fig. \ref{figR4b}, the oxide switch is used here as current-controlled NDR (Negative Differential Resistor) element. The simulation outputs in Fig. \ref{figR4c} correspond to the results originally published in \cite{pickett2012sub}.

\section{SPICE modeling of memcapacitive devices} \label{sec4}

\subsection{Model {\normalfont C.1:} Ideal memcapacitor} \label{secC1}

{\bf Model:} A voltage-controlled memcapacitor is defined by  \cite{diventra09a}
\begin{equation}
q=C\left( \phi(t) \right)V_\text{C}, \label{c1:eq1}
\end{equation}
where
\begin{equation}
\phi(t)=\int\limits_0^t V_\text{C}(\tau)\textnormal{d}\tau \label{c1:eq2}
\end{equation}
is the "flux". From application point of view, a memcapacitor switching between two limiting values of memcapacitance would be of value. This property is achieved, for example, in the following model resembling the memristor model given by Eq. (\ref{r1:eq6})
\begin{equation}
C(\phi(t))=C_\text{low}+\frac{C_\text{high}-C_\text{low}}{e^{-4k(\phi(t)+\phi_0)}+1}, \label{c1:eq3}
\end{equation}
where $C_\text{low}$ and $C_\text{high}$ are limiting values of memcapacitance ($C_\text{low}<C_\text{high}$), $k$ is a constant and $\phi_0$ is a constant defining the initial value of the capacitance $C_{ini}=C(\phi =0)$. In terms of the initial capacitance, Eq. (\ref{c1:eq3}) can be rewritten as follows:
\begin{equation}
C(\phi(t))=C_\text{low}+\frac{C_\text{high}-C_\text{low}}{ae^{-4k\phi (t)}+1},\;\;\; a=\frac{C_\text{high}-C_\text{ini}}{C_\text{ini}-C_\text{low}} \label{c1:eq4}
\end{equation}

{\bf Features:} The positive aspects of Eq. (\ref{c1:eq4}) model include its simplicity and switching between two limiting values. Among the negative ones we note a lack of
 switching threshold, sensitivity to fluctuations,  over-delayed switching \cite{diventra13b}, and a  possibility of active behavior \cite{diventra09a}.

\begin{figure}[tb]
 \begin{center}
    \includegraphics[width=7cm]{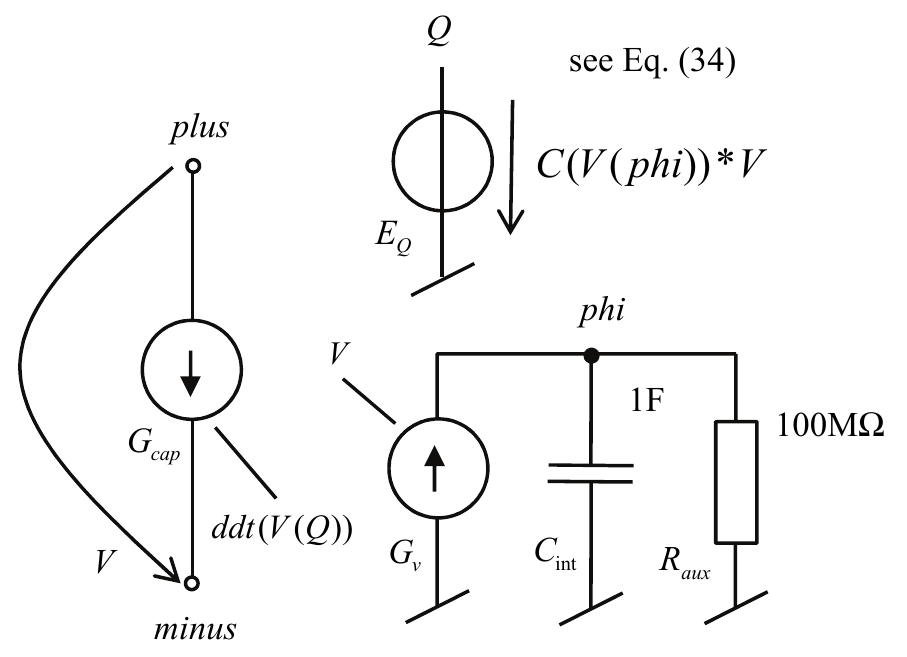}
\caption{\label{figC1a} Model of ideal memcapacitor from Section \ref{secC1}.}
\end{center}
\end{figure}

 The memcapacitor can be modeled as shown in Fig. \ref{figC1a}. The flux is computed as an integral of terminal voltage $V$: the controlled source $G_v$ whose current is equal to the voltage $V$ charges the capacitor $C_\text{int}$, thus the voltage of the node $phi$ is equal to the flux. This flux is then used to compute the memcapacitance according to Eq. (\ref{c1:eq4}). The charge is provided as a voltage of node $Q$ of the controlled voltage source $E_Q$. In such a way, the charge is available as a simulation result for inspection, without a necessity of its subsequent computation from the terminal current. The charge is then used for evaluating the terminal current via time-domain differentiation (see the source $G_\text{cap}$).

Note that in the simulation programs, which provide the feature of direct modeling of the charge sources (e.g. OrCAD PSpice v. 16, HSPICE, Micro-Cap), the source $G_\text{cap}$ can be implemented via this kind of source without the use of ddt operation (see the codes in Appendix \ref{app_C1}).
In case of need, the memcapacitive port can be also modeled as a parallel connection of a fixed capacitor $C_\text{low}$ and a variable capacitor according to Eq. (\ref{c1:eq4}).

\begin{figure}[tb]
 \begin{center}
    \includegraphics[width=9cm]{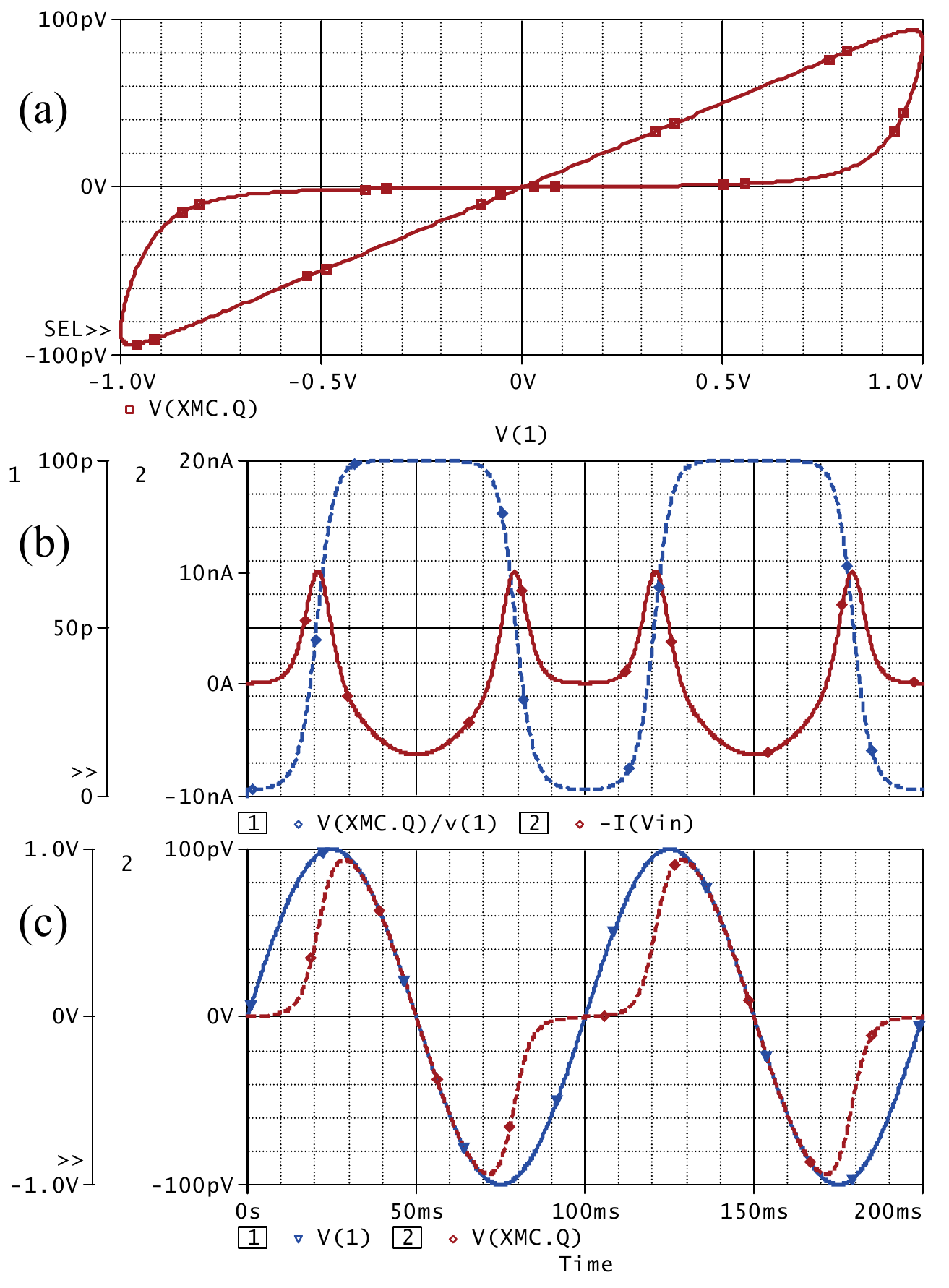}
\caption{\label{figC1b} Transient analysis of the ideal memcapacitor using Fig. \ref{figC1a} model: (a) pinched hysteresis loop, (b) memcapacitance (dashed blue line) and terminal current (solid red line), (c) terminal voltage (solid blue line) and charge (dashed red line).}
\end{center}
\end{figure}

{\bf  Results:} The subcircuit of ideal memcapacitor from Appendix \ref{app_C1}, based on the model from Fig. \ref{figC1a}, is used for simulating hard- switching phenomena which appear when exciting the memcapacitor with the parameters given in SPICE code of this subcircuit by 1V/1Hz sinusoidal voltage source. Figure \ref{figC1b} shows the PSpice outputs.

For checking the accuracy of the computation, several criteria can be used, for example the rule of the immediate steady state.
HSPICE provides the best results for Gear method and with the options RUNLVL=0 and LVLTIM=1.

\subsection{Model {\normalfont C.2:} Multilayer memcapacitive system} \label{secC2}

{\bf Model:} In a multilayer memcapacitive system, several metal layers are embedded into the dielectric medium separating capacitor plates \cite{martinez09a}. Here, we consider the simplest realization of such system involving two internal metal layers, which can be described as a first-order charge-controlled memcapacitive system \cite{martinez09a}:
\begin{eqnarray}
V_\text{C}&=&C^{-1}(Q,q)q \label{c2:eq1} \\
\frac{\textnormal{d}Q}{\textnormal{d}t}&=&I_{12}
\label{c2:eq2}
\end{eqnarray}
where
\begin{eqnarray}
C(Q,q)&=&\frac{C_0}{1+ \frac{\delta}{d} \frac{Q}{q}} \;\;, \label{c2:eq3}\\
I_{12}&=& \frac{S\,e}{2\pi
h\delta^2}\left[\left(U-\frac{eV_1}{2}\right)e^{-\frac{4\pi
\delta\sqrt{2m}}{h}\sqrt{U-\frac{eV_1}{2}}}\right.- \nonumber \\
 && \left.-\left(U+\frac{eV_1}{2}\right)e^{-\frac{4\pi
\delta_k\sqrt{2m}}{h}\sqrt{U+\frac{eV_1}{2}}}\right]
\label{c2:eq4}
\end{eqnarray}
if $eV_1<U$, and
\begin{eqnarray}
I_{12}&=&\frac{S\,e^3\,V_1^2}{4\pi
hU\delta^2}\left[e^{-\frac{4\pi
\delta\sqrt{m}U^{3/2}}{ehV_1}}\right. - \nonumber \\
&& \left. -\left(1+\frac{2eV_1}{U}\right)e^{-\frac{4\pi
\delta\sqrt{m}U^{3/2}}{ehV_1}\sqrt{1+\frac{2eV_1}{U}}}\right]
\label{c2:eq5} \;\;\;\;\;\;\;\;\;\;
\end{eqnarray}
if $eV_1>U$. Here,
\begin{equation}
 V_1=(q+Q)\delta/(S\varepsilon_0\varepsilon_\text{r}) \label{c2:eq6}
\end{equation}
is the voltage drop across internal layers, $Q$ is the internal layer charge, $S$ is the plate area, $d$ is the distance between plates, $\delta$ is the distance between internal layers placed symmetrically between the plates, $\varepsilon_0$ is the vacuum permittivity,
$\varepsilon_\text{r}$ is the relative dielectric constant of the insulating
material, $U$ is the potential barrier height between two
internal metal layers, $m$ and $e$ are electron mass and charge, respectively, $h$ is the Planck constant, and $C_0=\varepsilon_0
\varepsilon_\text{r} S/d$ is the capacitance of the system without internal metal layers. Note that Eqs. (\ref{c2:eq4}) and (\ref{c2:eq5}) are given for $V_1>0$. For $V_1<0$, the sign of $I_{12}$ should be changed and $|V_1|$ should be used in Eqs. (\ref{c2:eq4}), (\ref{c2:eq5}).

\begin{figure}[tb]
 \begin{center}
    \includegraphics[width=5cm]{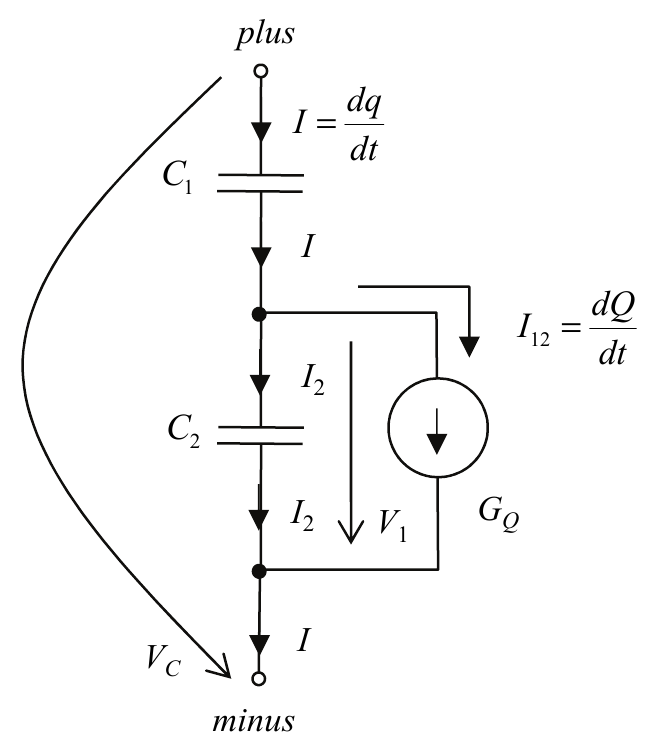}
\caption{\label{figC2a} Model of two-layer memcapacitive system described by Eqs. (\ref{c2:eq1})-(\ref{c2:eq6}).}
\end{center}
\end{figure}

{\bf Features:} Multilayer memcapacitive system is an example of memory device with the possibility of zero and negative response \cite{diventra13b}. As such, hysteresis curves of this device may not pass through the origin \cite{martinez09a,diventra13b}. It is shown in \cite{martinez09a} that the multilayer memcapacitive system can be modeled by an equivalent circuit, consisting of linear capacitors and nonlinear resistors. For the case of two layers, such circuit is modified to the form in Fig. 13, with nonlinear resistor modeled via a controlled current source $G_Q$. It can be shown that if capacitances $C_1$ and $C_2$ are set to values
\begin{equation}
C_1=\frac{\varepsilon_0 \varepsilon_\text{r} S}{d-\delta}=\frac{C_0}{1-\delta / d},\;\;\; C_2=\frac{\varepsilon_0 \varepsilon_\text{r} S}{\delta}=\frac{C_0}{\delta / d} \label{c2:eq7}
\end{equation}
and if the current flowing through the source $G_Q$ is $I_{12}$, given by Eqs. (\ref{c2:eq4}) and (\ref{c2:eq5}), then the circuit in Fig. \ref{figC2a} behaves as memcapacitive system with the memcapacitance given by Eq. (\ref{c2:eq3}), and that the voltage across $C_2$ is the voltage (\ref{c2:eq6}) across the internal layers. Then $I_2=I-I_{12}=\textnormal{d}(q-Q)/\textnormal{d}t$ and thus $C_2$ is charged to the charge $q-Q$. The voltage $V_1$ will be $(q-Q)/C_2$ which gives Eq. (\ref{c2:eq6}). The sum of voltages across $C_1$ and $C_2$ is $V_C = q/C(Q,q) = q/C_1+(q-Q)/C_2$. After substituting Eqs. (\ref{c2:eq7}) we get the formulae (\ref{c2:eq3}) for the memcapacitance.

The current $I_{12}$ from Eqs. (\ref{c2:eq4}) and (\ref{c2:eq5}), representing formulae for current-voltage characteristic of electric tunnel junction, takes the values from a large dynamic range which exceeds the numerical limits of SPICE-family simulation programs. For typical numerical values given in Appendix \ref{app_C2}, $I_{12}$ is of about $10^{-127}$ for $eV_1/U=0.1$, $10^{-116}$ for $eV_1/U=1$, $10^{-56}$ for $eV_1/U=2$, and $10^{-6}$ for $eV_1/U=10$. It turns out that the low-voltage range $eV_1<U$ (Eq. \ref{c2:eq4}) generates the currents much below the numerical threshold of SPICE, and that the first term of Eq. (\ref{c2:eq5}) approximates well the $I_{12}$ versus $V_1$ dependence in the form
\begin{equation}
I_{12}\approx aV_1 |V_1|e^{-\frac{b}{|V_1|}}, \;\;\; a=\frac{Se^3}{4\pi h U \delta^2}, \;\;\; b=\frac{4\pi\delta \sqrt{m}U^{\frac{3}{2}}}{eh} \label{c2:eq8}
\end{equation}
both for positive and negative values of $V_1$. The SPICE codes presented below can be easily modified to include the second term of Eq. (\ref{c2:eq5}) if required.

To prevent numerical underflow, it is useful to compute logarithm of $I_{12}$ from (\ref{c2:eq8}), to limit artificially its range, and then to compute $I_{12}$ via inverse logarithm from this limited values. Examples of the corresponding SPICE codes, providing reliable computation, are given in Appendix \ref{app_C2}.
Note that due to undocumented errors in OrCAD PSpice v. 16 and HSPICE, they handle incorrectly numerical parameters which underflow the limit of ca $10^{-30}$. In this model, such parameters are electron mass $m$ and Planck constant $h$. That is why the codes for PSpice and HSPICE are modified accordingly for computing auxiliary variables $a$ and $b$ from (\ref{c2:eq8}) which depend on these quantities.

\begin{figure}[tb]
 \begin{center}
    \includegraphics[width=9cm]{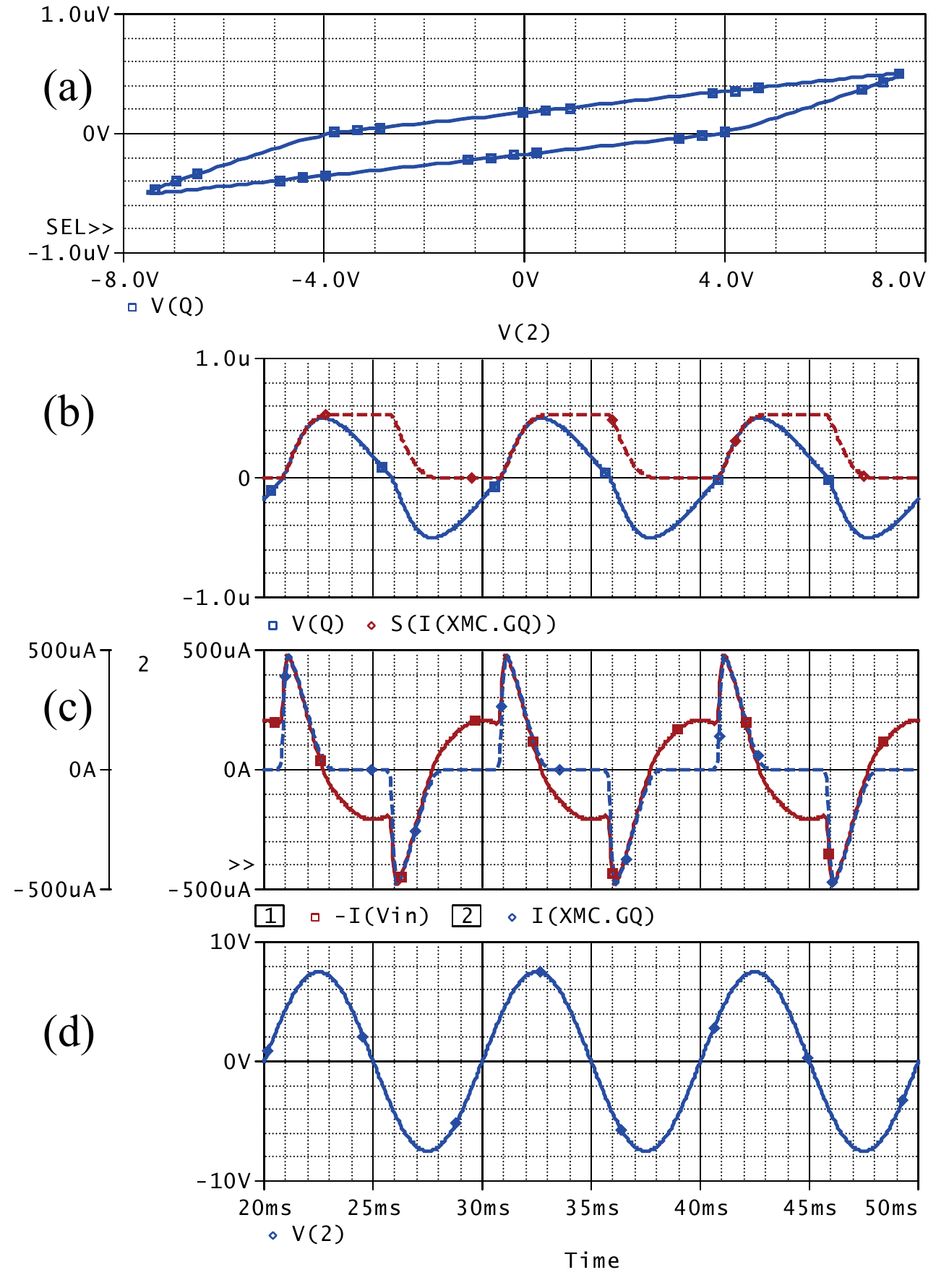}
\caption{\label{figC2b} Transient analysis of memcapacitive system from Fig. \ref{figC2a} which is driven by sinusoidal 7.5V/100Hz voltage source with 1Ohm serial resistance, (a) charge-voltage hysteresis loop, (b) memcapacitor charge (solid blue) and charge of the internal layers (dashed red), (c) terminal current (solid red) and current $I_{12}$ (dashed blue), (d) exciting voltage.}
\end{center}
\end{figure}

{\bf Results:} The SPICE codes from Appendix \ref{app_C2} can provide all the simulation results from \cite{martinez09a}. Figure \ref{figC2b} confirms the nonpinched charge-voltage hysteresis loop. This model enables studying all the interesting phenomena described in \cite{martinez09a}, including frequency dependent hysteresis, diverging and negative capacitance.

\subsection{Model {\normalfont C.3:} Bistable membrane memcapacitive system} \label{secC3}

{\bf Model:} The model of bistable membrane memcapacitive system \cite{pershin11c} is specified by
\begin{eqnarray}
q(t)&=& C(y) V(t),  \label{c3:eq1}\\
\frac{\textnormal{d}y}{\textnormal{d}\tau}&=&\dot{y}, \label{c3:eq2}\\
\frac{\textnormal{d}\dot{y}}{\textnormal{d}\tau}&=&
-4\pi^2\,y\,\left(\left(\frac{y}{y_0}\right)^2-1\right)-\Gamma\,\dot{y}-\left(\frac{\beta(\tau)}{1+y}\right)^2, \;\;\;\; \label{c3:eq3}
\end{eqnarray}
where
\begin{equation}
C(y)=\frac{C_0}{1+y}, \label{c3:eq4}
\end{equation}
$y_0=z_0/d$, $\Gamma=2\pi\,\gamma/\omega_0$, $\beta(t)=\left[2\pi
/ \left( \omega_0\,d \right) \right] \sqrt{C_0 / \left( 2\,m\right)
}\,V(t)$ and time derivatives are taken with respect to the
dimensionless time $\tau=t\,\omega_0 / \left( 2\pi \right)$. Here,
$\pm z_0$ are the equilibrium positions of the membrane,
$d$ is the separation between the bottom plate and middle position of the
flexible membrane, $\gamma$ is the damping
constant, $\omega_0$ is the natural angular frequency of the
system, $m$ is the mass of the membrane and $C_0=\epsilon_0\,S/d$.
The dimensionless membrane displacement $y$ and membrane's velocity
$\dot{y}$ play the role of the internal state variables.

\begin{figure}[tb]
 \begin{center}
    \includegraphics[width=8cm]{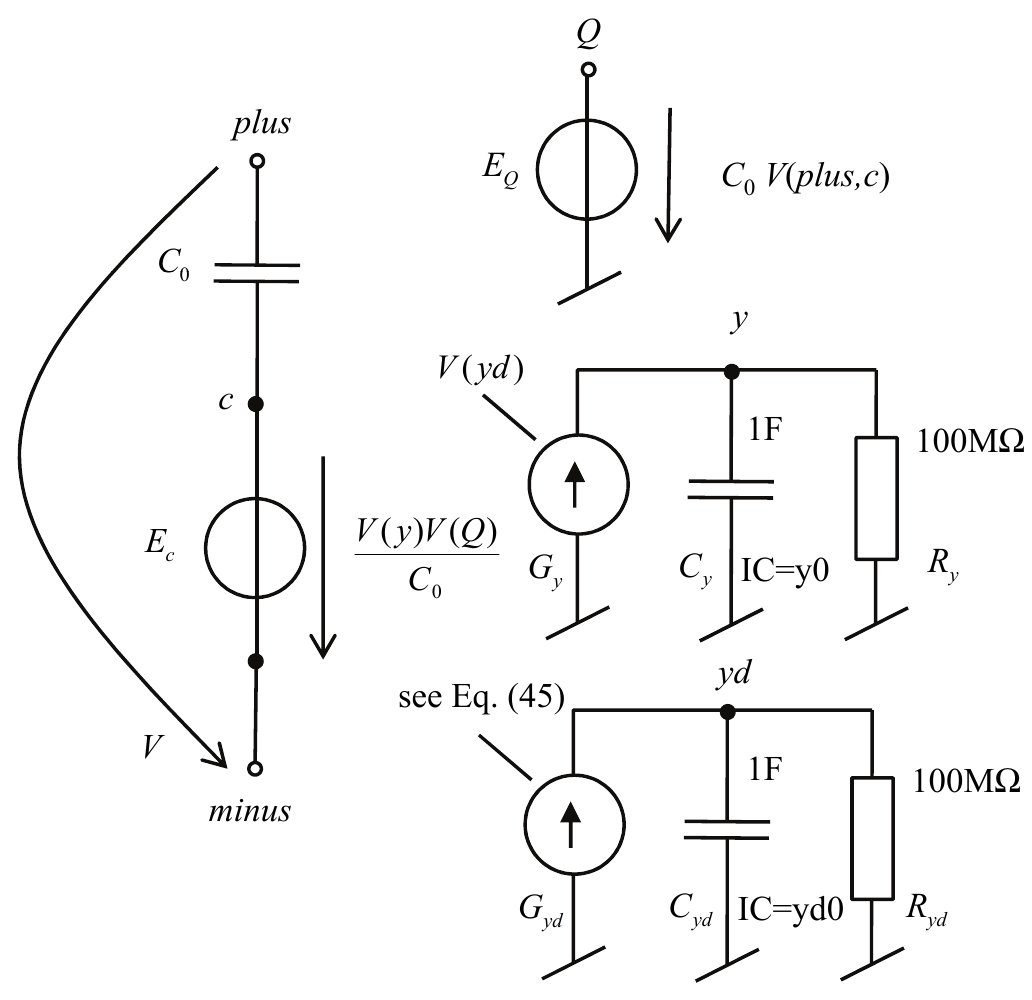}
\caption{\label{figC3a} Model of bistable membrane memcapacitive system described by Eqs. (\ref{c3:eq1})-(\ref{c3:eq4}).}
\end{center}
\end{figure}

{\bf Features:} This model describes a memcapacitive device with two well-defined equilibrium states ideally suited for binary applications.
In order to model reliably the memcapacitive port, Eqs. (\ref{c3:eq1}) and (\ref{c3:eq4}) are arranged to the form
\begin{equation}
V(t)=\frac{1}{C_0}q(t)+\frac{y}{C_0}q(t),
\label{c3:eq5}
\end{equation}
which represents the serial connection of two capacitors, with fixed capacitance $C_0$ and with the capacitance dependent on the variable $y$ (the fact that $y$ can take negative values does not cause any problems). The second one is modeled in Fig. \ref{figC3a} via a controlled voltage source $E_c$. The charge, which is necessary for computing the source voltage, can be obtained by integrating the terminal current, or more conveniently, it is directly the product of voltage across the capacitor $C_0$ and its capacitance. The charge value is available as a voltage of the voltage source $E_Q$. Two classical integrator circuits provide the computation of $y$ and $\dot{y}$ quantities according to Eqs. (\ref{c3:eq2}) and (\ref{c3:eq3}), representing them as voltages of nodes $y$ and $yd$. Surprisingly, HSPICE provides low precision of the simulated waveforms with this model. The precision is considerably increased after modeling the variable part of the memcapacitive port directly by a capacitor with formula-controlled capacitance (see the Appendix \ref{app_C3}).

\begin{figure}[tb]
 \begin{center}
    \includegraphics[width=9cm]{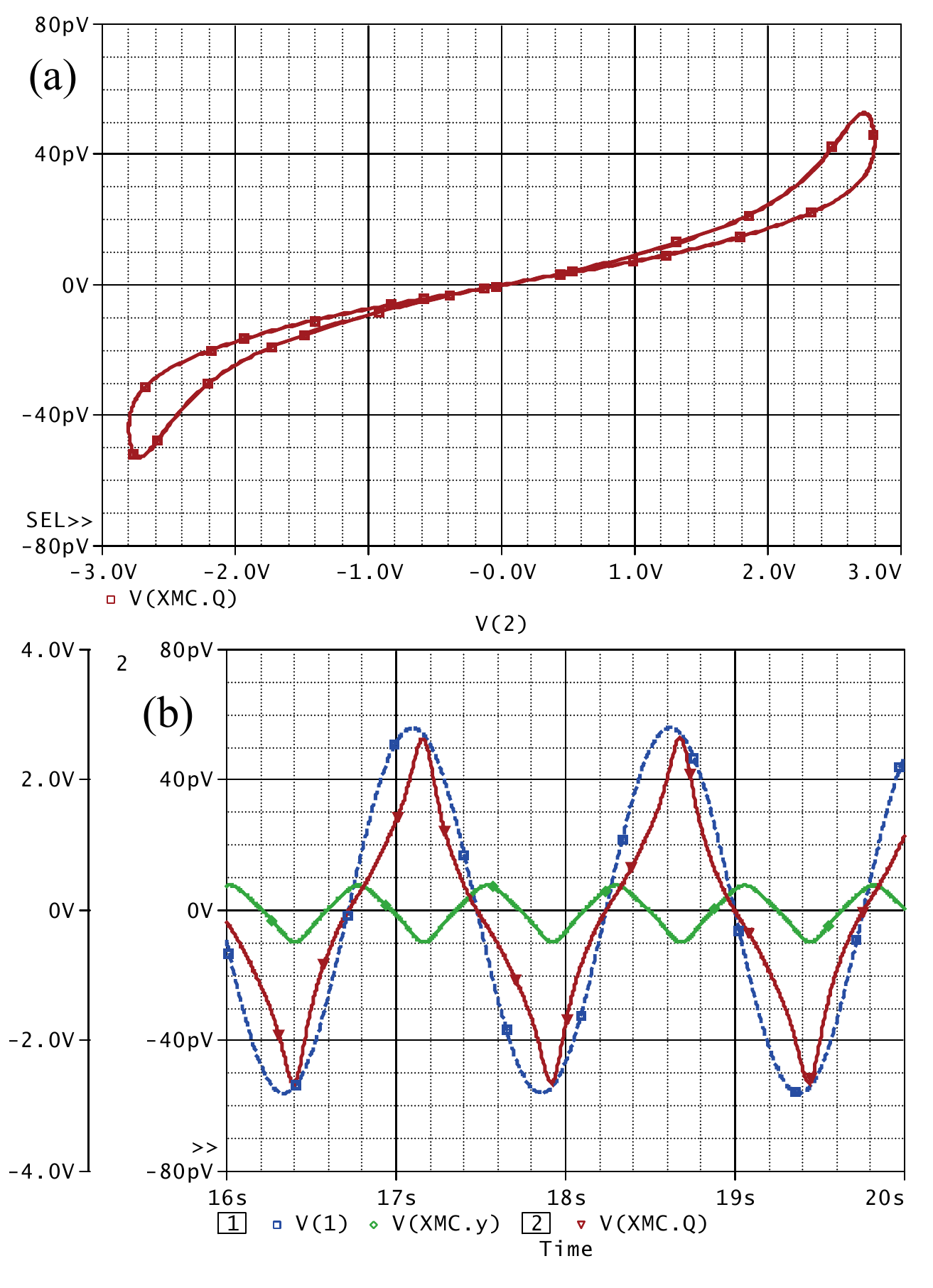}
\caption{\label{figC3b} Transient analysis of bistable elastic memcapacitive system from Fig. \ref{figC3a} in the periodical steady state under conditions defined in SPICE codes in Appendix \ref{app_C3}: (a) charge-voltage pinched hysteresis loop, (b) exciting sinusoidal voltage (dashed blue line), membrane position $y$ (green line), memcapacitor charge (red line).}
\end{center}
\end{figure}

{\bf Results:} Figure \ref{figC3b} shows some outputs of PSpice transient analysis of bistable memcapacitive device under the sinusoidal excitation. The simulation model confirms all the phenomena which are analyzed in Ref. \cite{pershin11c}, including the fact that the hysteresis is seen at intermediate frequencies compared to the natural frequency of the system. This model also offers the ability to analyze the dynamics of membrane under the voltage pulse excitation. In addition, the chaotic behavior of the device can be observed under the conditions specified in \cite{pershin11c}.

\subsection{Model {\normalfont C.4:} Bipolar memcapacitive system with threshold} \label{secC4}

{\bf Model:} Here we consider a generic model of memcapacitive devices with threshold. This model is formulated similarly to  the model of memristive device with threshold \ref{secR2} proved to be useful in many cases.  We assume that the memcapacitance $C$ plays the role of the internal state variable $x$, namely, $x \equiv C$, defining the device state via the following equations
\begin{eqnarray}
q&=&xV_\text{C}, \label{c4:eq1} \\
\frac{\textnormal{d}x}{\textnormal{d}t}&=&f(V_\text{C})W(x,V_\text{C}) \label{c4:eq1a}
\end{eqnarray}
where $f(.)$ is a function modeling the device threshold property (see Fig. \ref{figR2a}) and $W(.)$ is a window function:
\begin{eqnarray}
f(V_\text{C})&=&\beta \left( V_\text{C}-0.5\left[ |V_\text{C}+V_\text{t}|-|V_\text{C}-V_\text{t}| \right]\right) ,\label{c4:eq1b} \\
W(x,V_\text{C})&=&\theta\left( V_\text{C}\right) \theta\left(
C_\text{high}-x\right)+ \theta\left(- V_\text{C}\right) \theta\left(
x-C_\text{low}\right) . \;\;\;\;\; \label{c4:eq2}
\end{eqnarray}
Here $\theta(\cdot)$ is the step function, $\beta$ is a positive constant
characterizing the rate of memcapacitance change when $|V_\text{C}|> V_\text{t}$,
$V_\text{t}$ is the threshold voltage, and  $C_\text{low}$ and $C_\text{high}$ are limiting
values of the memcapacitance $C$. In Eq. (\ref{c4:eq2}), the role of $\theta$-functions
is to confine the memcapacitance change to the interval between
$C_\text{low}$ and $C_\text{high}$.

\begin{figure}[tb]
 \begin{center}
    \includegraphics[width=8cm]{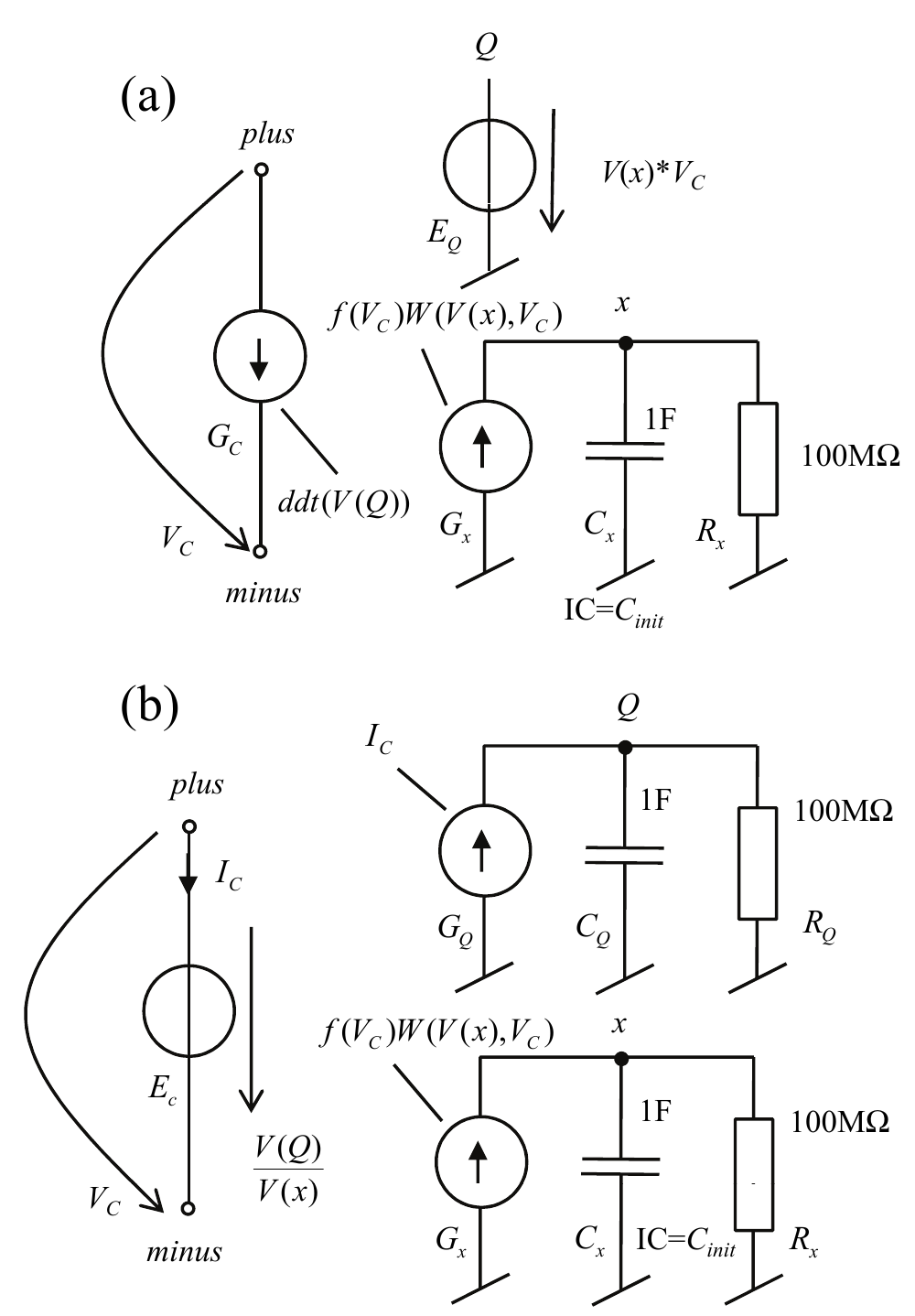}
\caption{\label{figC4a} Two equivalent models of the memcapacitive device with threshold.}
\end{center}
\end{figure}

{\bf Features:} The threshold property is not only a widespread attribute of many physical devices but also an attractive feature from the application point of view. While the present model is formulated without keeping any specific memcapacitive device in mind, its structure is closely related to the model of bipolar memristive devices with threshold and thus can describe a memcapacitive component of such devices, which might be the major one in properly designed structures. The positive aspects of the present model include the existence of the switching threshold and limiting values of memcapacitance. We note, however, that such a model may, in some cases, result in an active device behavior.

Fig. \ref{figC4a} shows two possible models based on Eqs. (\ref{c4:eq1})-(\ref{c4:eq2}). Both of them compute uniformly the state variable $x$ via integrating Eq. (\ref{c4:eq1a}) (see $G_x$, $C_x$, $R_x$). In the model (a), charge is computed as a product of memcapacitor voltage $V_C$ and memcapacitance which is resented by the voltage $V(x)$ (see the controlled source $E_Q$). The memcapacitor current, i.e. time derivative of the charge, is provided by the controlled current source $G_C$. The model (b) avoids the differentiation: the charge is computed via integrating the current $I_C$ flowing through the memcapacitive port, and the terminal voltage is computed as a ratio of the charge and capacitance. The division by $V(x)$ is not dangerous since the denominator is changing within the limits from $C_\text{low}$ to $C_\text{high}$.

\begin{figure}[tb]
 \begin{center}
    \includegraphics[width=8cm]{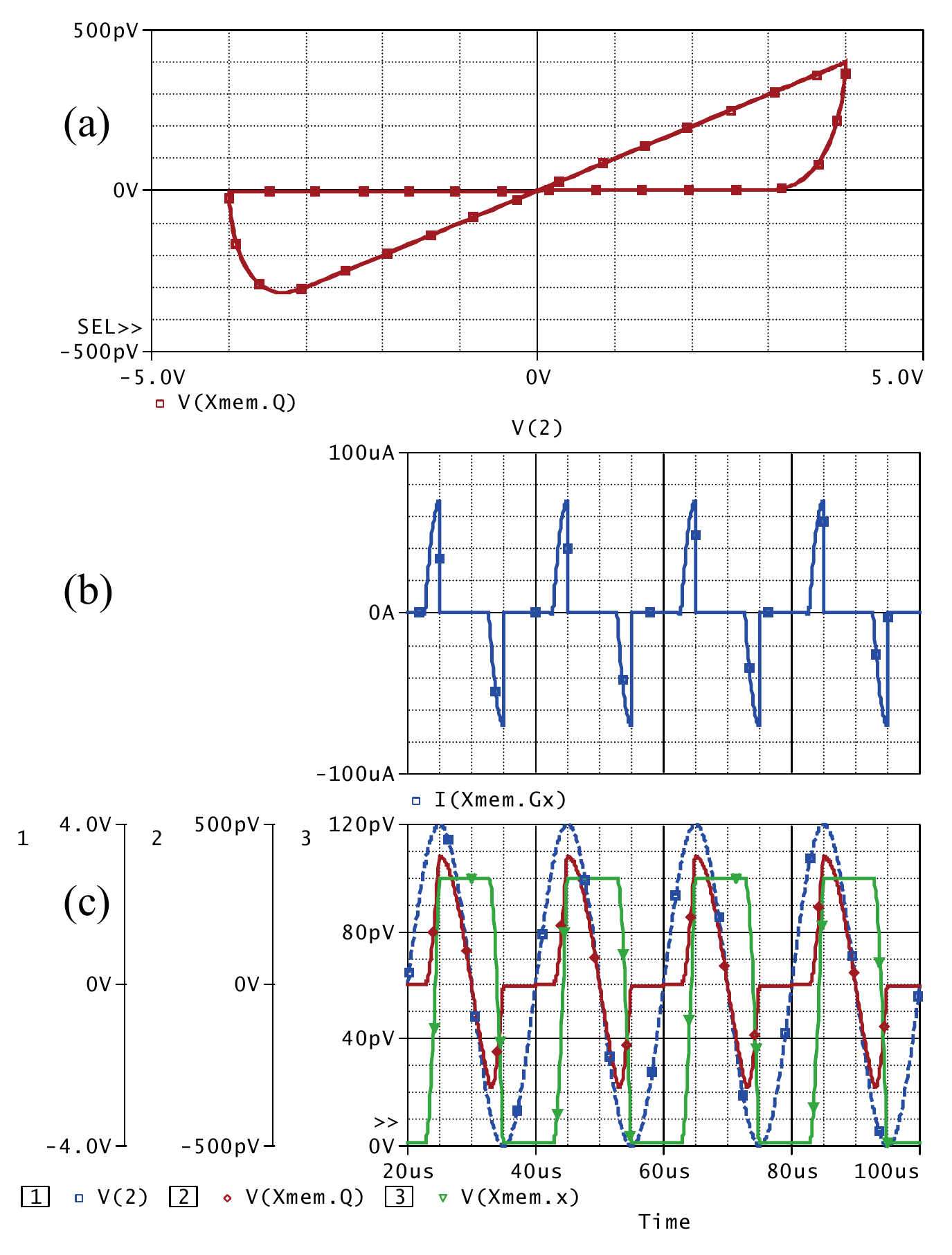}
\caption{\label{figC4b} Transient analysis of the model from Fig. \ref{figC4a}(b). Memcapacitive device with threshold voltage $V_t=3$V is driven by sinusoidal 4V/50kHz signal: (a) charge-voltage pinched hysteresis loop, (b) time derivative of the memcapacitance (i.e. current charging $C_x$ in Fig. \ref{figC4a}), (c) exciting voltage (blue dashed line), memcapacitor charge (red line), memcapacitance (green line).}
\end{center}
\end{figure}

Both models provide good results in PSpice and LTspice. However, simulations in HSPICE are accompanied by serious accuracy (model (a)) and convergence (model (b)) problems. Their nature probably consists in undocumented problems in HSPICE Version A-2008.03. They can be overcome via running the HSPICE-RF simulator from the software package instead of HSPICE. The Appendix \ref{app_C4} provides PSpice and LTspice codes based on the model in Fig. \ref{figC4a} (b), and HSPICE code for the same model which can be run on HSPICE-RF. Fig. \ref{figC4b} shows the simulation results from PSpice, demonstrating the periodical switching of the memcapacitance between $C_\text{low}$ and $C_\text{high}$  states.

%Possible parameters: $C_{low}=1$pF, $C_{high}=100$pF, $V_t=3$V, $\beta=10^7$pF/(V s). These parameters assume $\mu$s switching times.

\section{SPICE modeling of meminductive devices} \label{sec5}

\subsection{Model {\normalfont L.1:} Ideal meminductor} \label{secL1}

\begin{figure}[tb]
 \begin{center}
    \includegraphics[width=7cm]{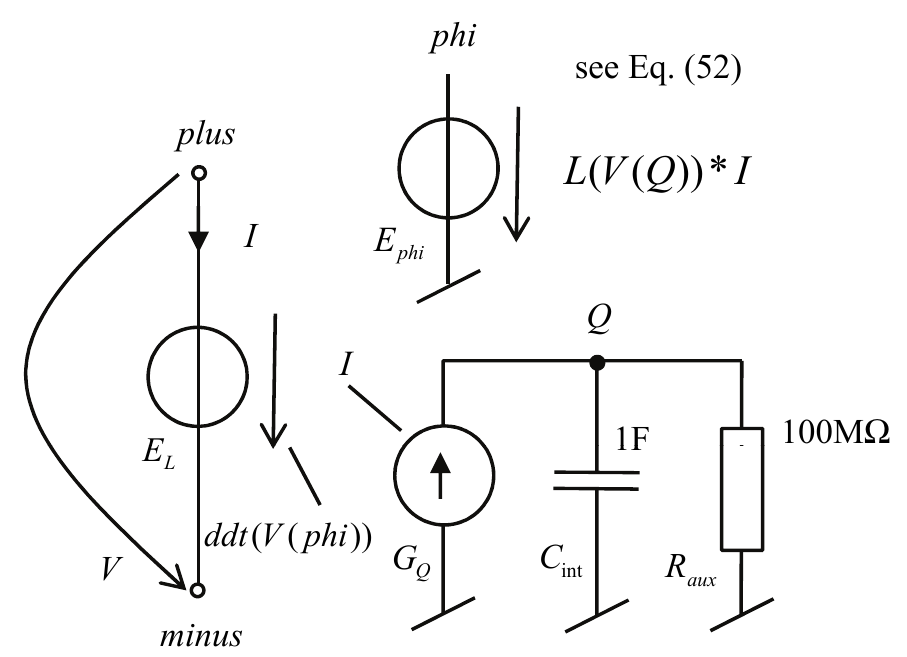}
\caption{\label{figL1a} Ideal meminductor implementation in SPICE.}
\end{center}
\end{figure}

{\bf Model:} A current-controlled meminductor is defined as  \cite{diventra09a}
\begin{equation}
\phi=L( q(t)) I, \label{l1:eq1}
\end{equation}
where the charge $q(t)$ is the integral of the current. From application point of view, a meminductor switching between two limiting values of meminductance is desirable.
Similarly to Eqs. (\ref{r1:eq6}) and (\ref{c1:eq3}), we formulate a model of such meminductor as
\begin{equation}
L(q(t))=L_\text{low}+\frac{L_\text{high}-L_\text{low}}{e^{-4k(q(t)+q_0)}+1}, \label{l1:eq2}
\end{equation}
where $L_\text{low}$ and $L_\text{high}$ are limiting values of meminductance ($L_\text{low}<L_\text{high}$). The meminductance can be derived also as a function of the initial inductance $L_\text{ini}=L(q=0)$:
\begin{equation}
L(q(t))=L_\text{low}+\frac{L_\text{high}-L_\text{low}}{ae^{-4kq(t)}+1},\;\;\; a=\frac{L_\text{high}-L_\text{ini}}{L_\text{ini}-L_\text{low}} \label{l1:eq3}
\end{equation}

{\bf Features:} Positive aspects of Eq. (\ref{l1:eq3}) model include its simplicity and switching between two limiting values. Among the negative ones we note a lack of
 switching threshold, sensitivity to fluctuations,  over-delayed switching \cite{diventra13b}, and the  possibility of active behavior \cite{diventra09a}.
 The meminductor can be modeled in a similar way as the memcapacitor from Section \ref{secC1}, see Fig. \ref{figL1a}. The port current $I$ is integrated into the voltage of node $Q$, representing the charge. According to Eqs. (\ref{l1:eq1}) and (\ref{l1:eq3}), the flux is evaluated as the voltage of the controlled voltage source $E_{phi}$. This voltage is then used for computing the terminal voltage via time-domain differentiation (see the source $E_L$).
Note that in the simulation programs, which provide the feature of direct modeling of the flux sources (e.g. OrCAD PSpice v. 16, HSPICE, Micro-Cap), the source $E_L$ can be implemented via this kind of source without the use of ddt operation (see the codes in Appendix \ref{app_L1}).
If necessary, the meminductive port can be also modeled as a serial connection of a fixed inductor $L_\text{low}$ and a variable inductor according to Eq. (\ref{l1:eq3}).

\begin{figure}[tb]
 \begin{center}
    \includegraphics[width=9cm]{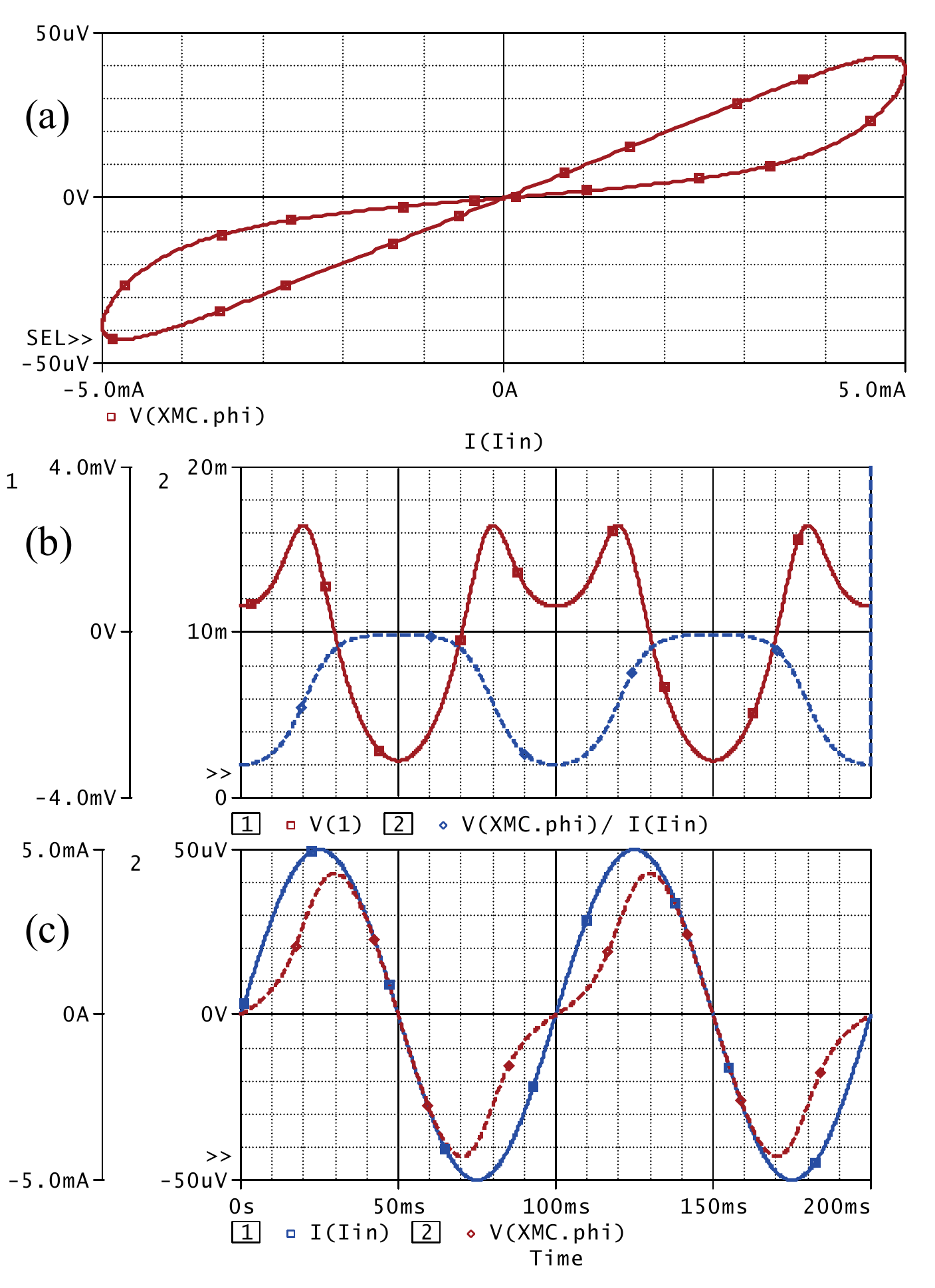}
\caption{\label{figL1b} Transient analysis of meminductor from Fig. \ref{figL1a}: (a) pinched hysteresis loop, (b) meminductance (dashed blue line) and terminal voltage (solid red line), (c) terminal current (solid blue line) and flux (dashed red line).}
\end{center}
\end{figure}

{\bf Results:} Results of the transient analysis in PSpice in Fig. \ref{figL1b} were obtained from the code in Appendix \ref{app_L1}. The meminductor is driven by the ideal current source, generating sinusoidal 5mA/10Hz waveform. The simulation results exhibit all basic fingerprints of the meminductor, i.e. odd-symmetric flux-current pinched hysteresis loop and its high-frequency shrinking property, unambiguous meminductance-charge state map, etc.

\subsection{Model {\normalfont L.2:} Effective meminductive system} \label{secL2}

\begin{figure}[tb]
 \begin{center}
    \includegraphics[width=5cm]{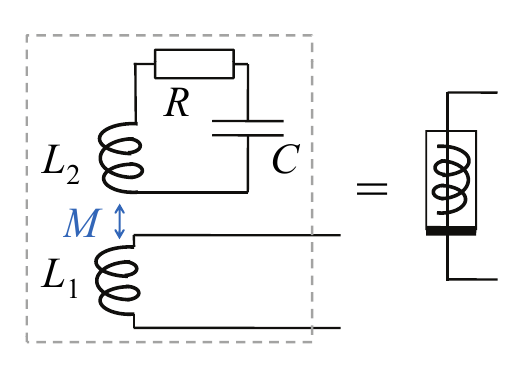}
\caption{\label{figL2a} Flux-controlled meminductive system based on the inductive coupling of a coil $L_1$ with a LCR contour \cite{Cohen12a}. Here, the
mutual inductance $M$ is equal to $k\sqrt{L_1L_2}$, where $0 \leq k \leq 1$ is the coupling coefficient.}
\end{center}
\end{figure}

{\bf Model:} Let us consider an example of an effective meminductive system that can be realized using
traditional circuit elements \cite{Cohen12a}. It consists of an
LCR contour inductively coupled to an inductor as shown in Fig. \ref{figL2a}. In this scheme, two inductors $L_1$ and $L_2$
interact with each other magnetically. The charge on the capacitor $C$ and the current through the inductor
$L_2$ play the role of internal state variables, namely, $x_1=q_C$ and $x_2=I_{L_2}$. The system is described by
\begin{eqnarray}
\phi &=& L\left( x_2,I\right) I, \label{l2:eq1}  \\
\frac{\textnormal{d}x_1}{\textnormal{d} t} &=& -x_2, \label{l2:eq2} \\
\frac{\textnormal{d}x_2}{\textnormal{d} t} &=& \frac{1}{L_2}\left( \frac{x_1}{C}-Rx_2 -M \frac{\textnormal{d}I}{\textnormal{d} t} \right), \label{l2:eq3}
\end{eqnarray}
where
\begin{equation}
L\left( x_2,I\right)=L_1+M\frac{x_2}{I} \;\; . \label{l2:eq4}
\end{equation}

\begin{figure}[tbp]
 \begin{center}
    \includegraphics[width=8cm]{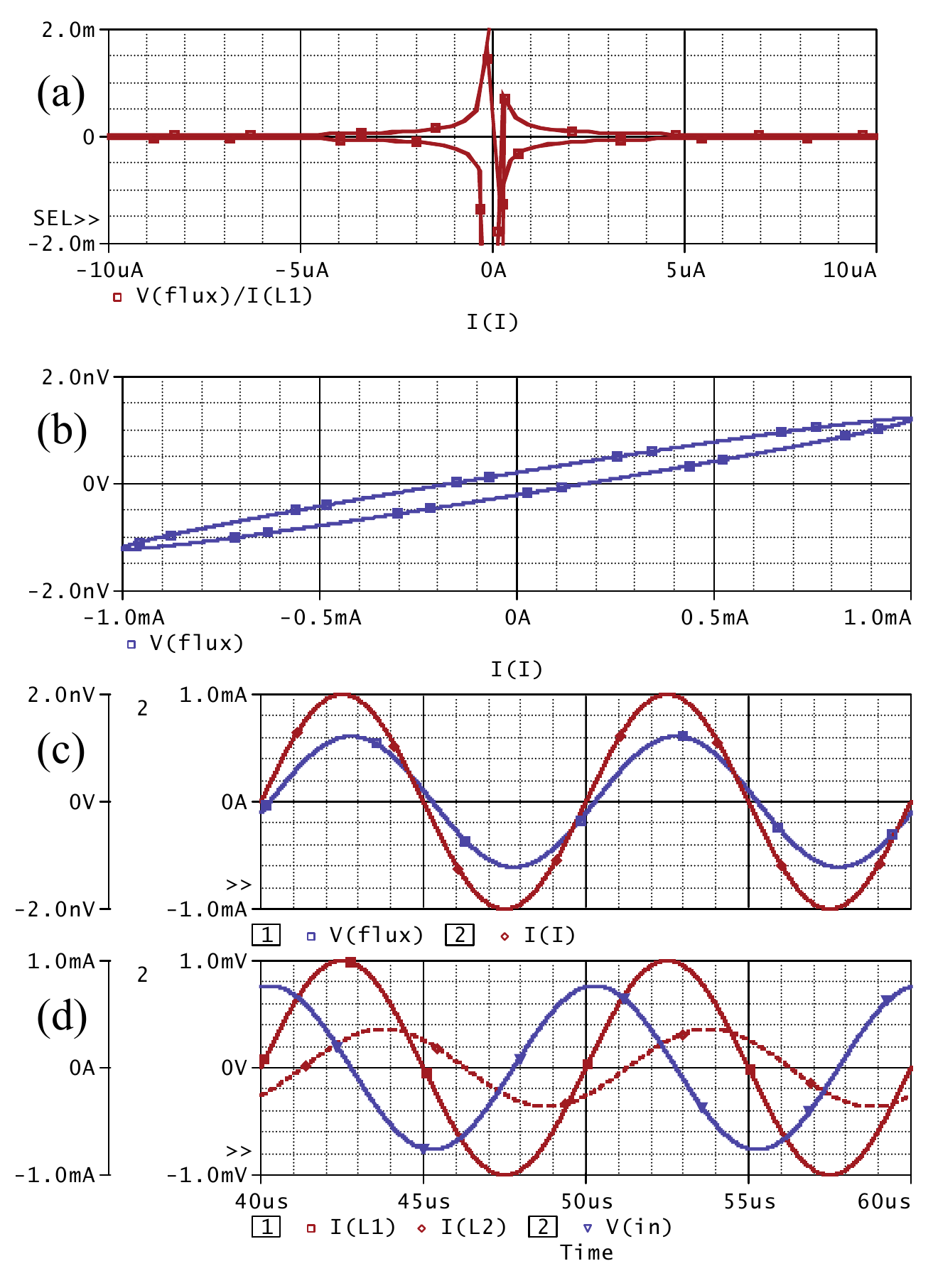}
\caption{\label{figL2b} Steady-state transient analysis of system from Fig. \ref{figL2a}: (a) Meminductance vs. current according to Eq. (\ref{l2:eq4}), (b) nonpinched flux-current hysteresis loop, (c) flux (red line) and terminal current (blue line), (d) currents through $L_1$ (solid red line) and $L_2$ (dashed red line) and voltage across $L_1$ (blue line).}
\end{center}
\end{figure}

{\bf Features:}  The circuit sketched in Fig. \ref{figL2a} does not offer non-volatile information storage and has certain similarities with elastic memcapacitive systems \cite{pershin11a}. Additional functionalities open if the resistor or capacitor (or both) is replaced by a memristive or memcapacitive system, respectively. We note that although $\textnormal{d}I/ \textnormal{d} t$ enters into the right-hand-side of Eq. (\ref{l2:eq3}) and thus this equation differs from the canonical equations of meminductive systems \cite{diventra09a}, such a definition of meminductive system can be considered as a reduced (effective) one \cite{traversa13a} that can be written in the canonical form using additional internal state variables \cite{traversa13a}.

{\bf Results:} Since the circuit in Fig. \ref{figL2a} is a simple linear dynamical system containing conventional passive elements, its SPICE modeling does not require any special approach. Figure \ref{figL2b} shows some results of PSpice transient analysis. The inductor $L_1$ is driven by a current source with sinusoidal 1mA/100kHz waveform. Since $x_2$ = $I(L_2)$, the effective meminductance according to Eq. (\ref{l2:eq4}) depends on the ratio of currents flowing through inductors $L_2$ and $L_1$. It is obvious from the waveforms $I(L_1)$ and $I(L_2)$ in Fig. \ref{figL2b} (d) that the meminductance can take infinite values when $I(L_1)$ crosses zero level and that both positive and negative values can be possible. This fact is confirmed in Fig. \ref{figL2b} (a). Figure \ref{figL2b} (c) shows that, due to the frequency dependent phase shift between flux and exciting current, the zero-crossing points of these waveforms are not identical, and thus the flux-current hysteresis loop in Fig. \ref{figL2b} (b) cannot be pinched.

\subsection{Model {\normalfont L.3:} Bipolar meminductive system with threshold} \label{secL3}

{\bf Model:} Here we consider a generic model of meminductive devices with current threshold. This model is formulated similarly to  the model of memristive device with threshold proved to be useful in many cases.  We assume that the meminductance $L$ plays the role of the internal state variable $x$, namely, $x \equiv L$, defining the device state via the following equations
\begin{eqnarray}
\phi&=&LI, \label{l3:eq1} \\
\frac{\textnormal{d}x}{\textnormal{d}t}&=&f(I)W(x,I) \label{l3:eq1a}
\end{eqnarray}
where $f(.)$ is a function modeling the device threshold property (see Fig. \ref{figR2a}) and $W(.)$ is a window function:
\begin{eqnarray}
f(I)&=&\beta \left( I-0.5\left[ |I+I_\text{t}|-|I-I_\text{t}| \right]\right) ,\label{l3:eq1b} \\
W(x,I)&=&\theta\left( I\right) \theta\left(
L_\text{high}-x\right)+ \theta\left(- I\right) \theta\left(
x-L_\text{low}\right) . \;\;\;\;\; \label{l3:eq2}
\end{eqnarray}
Here $\theta(\cdot)$ is the step function, $\beta$ is a positive constant
characterizing the rate of meminductance change when $|I|> I_\text{t}$,
$I_\text{t}$ is the threshold current, and  $L_\text{low}$ and $L_\text{high}$ are limiting
values of the meminductance $L$. In Eq. (\ref{l3:eq2}), the role of $\theta$-functions
is to confine the meminductance change to the interval between
$L_\text{low}$ and $L_\text{high}$.

\begin{figure}[tb]
 \begin{center}
    \includegraphics[width=8cm]{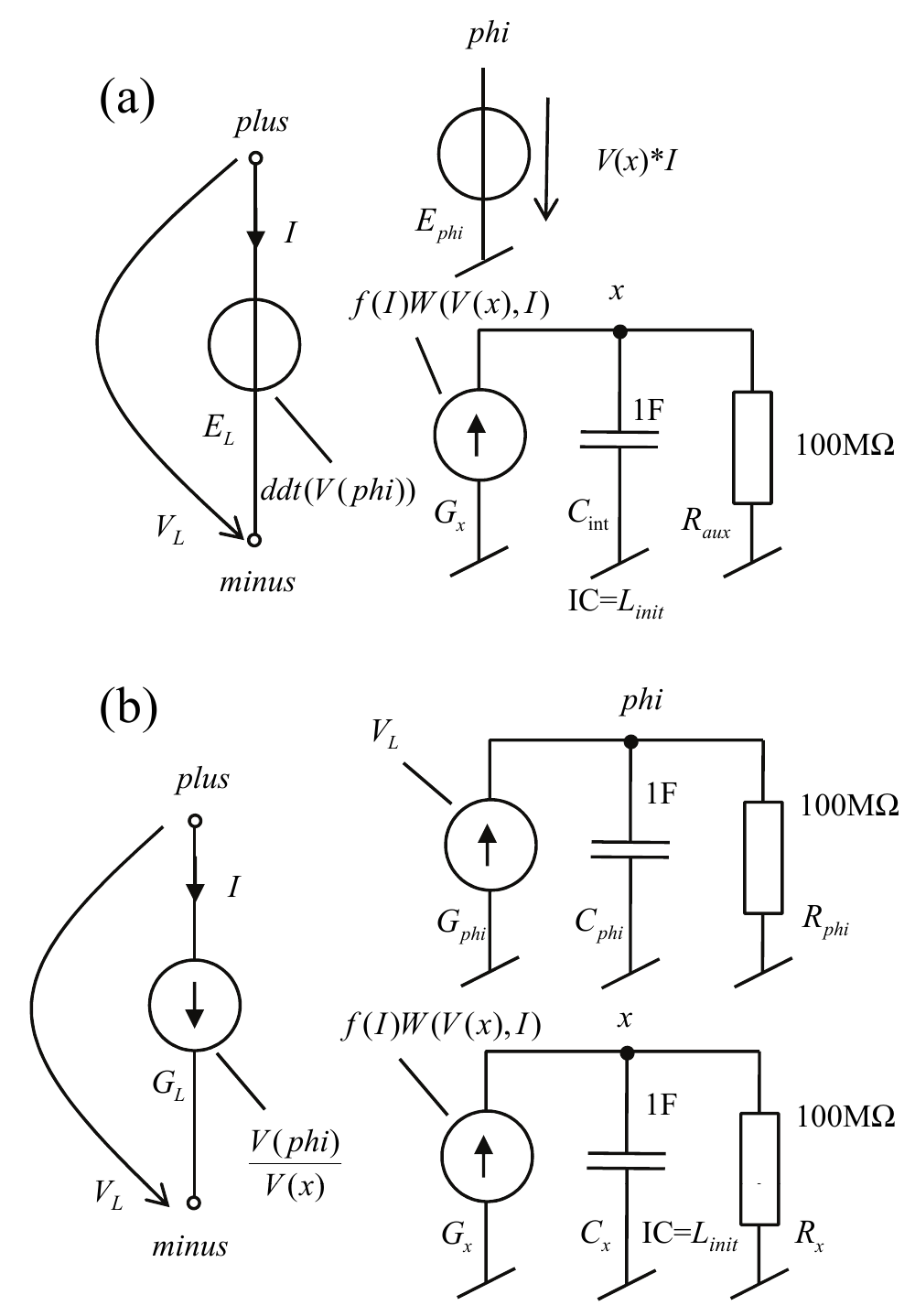}
\caption{\label{figL3a} Two equivalent models of the meminductive device with threshold.}
\end{center}
\end{figure}

{\bf Features:} The threshold property is not only a widespread attribute of many physical devices but also an attractive feature from the application point of view. The present model, however, is formulated without keeping any specific meminductive device in mind. The positive aspects of this model include the existence of the switching threshold and limiting values of meminductance. We note, however, that such a model may, in some cases, result in an active device behavior.

\begin{figure}[tb]
 \begin{center}
    \includegraphics[width=8cm]{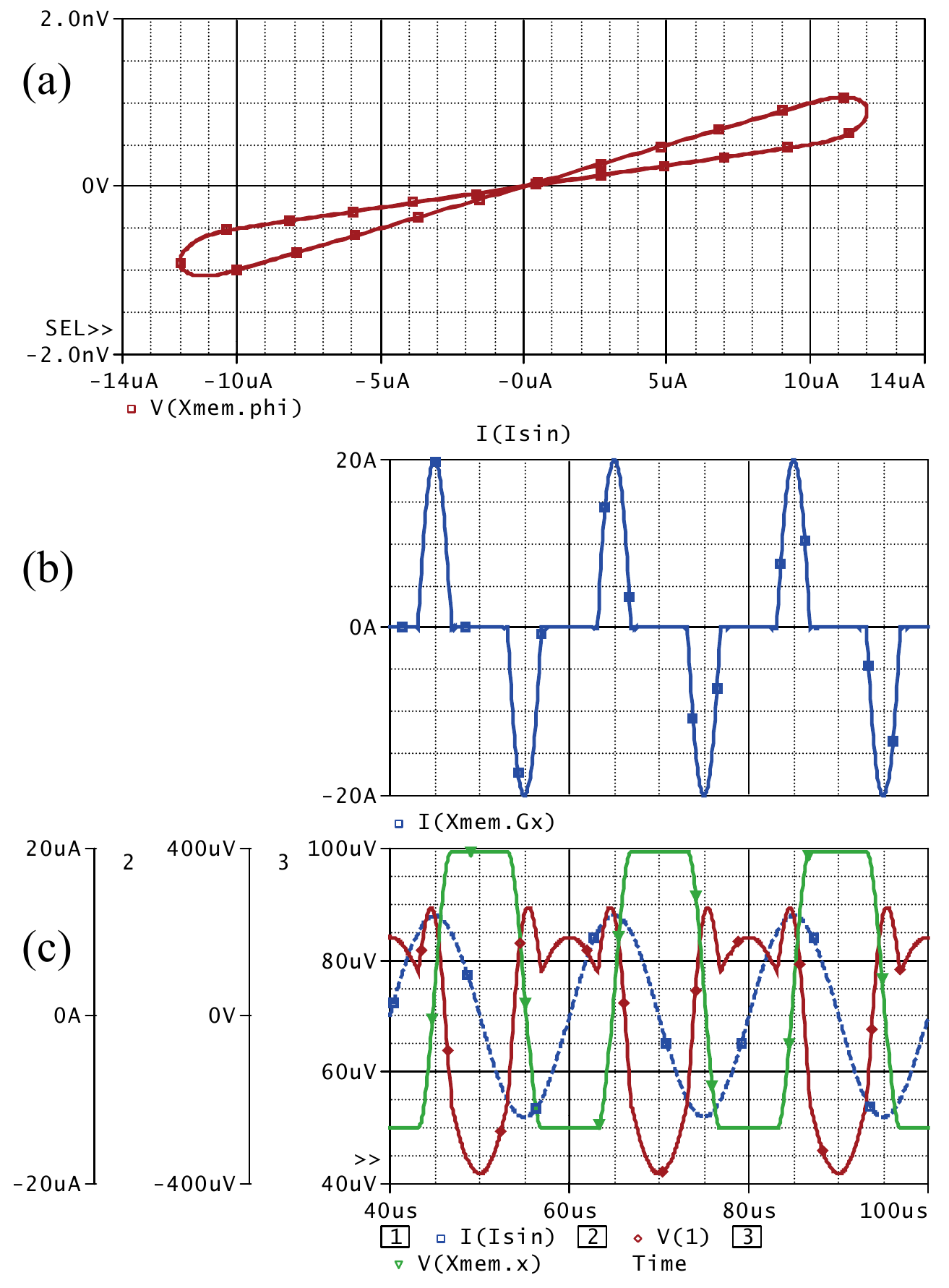}
\caption{\label{figL3b} Transient analysis of the model from Fig. \ref{figL3a}(a). Meminductive device with threshold current $I_t=10\mu$A is driven by sinusoidal $12\mu$A/50kHz signal: (a) flux-current pinched hysteresis loop, (b) time derivative of the meminductance (i.e. current charging $C_x$ in Fig. \ref{figL3a}), (c) exciting current (blue dashed line), meminductor voltage (red line), meminductance (green line).}
\end{center}
\end{figure}

Two kinds of SPICE-oriented models of the meminductive system with threshold are shown in Fig. \ref{figL3a}(a) and (b). In both cases, the state variable $x$, denoting the meminductance, is represented by the voltage of the node $x$ which is computed via time-domain integration according to Eq. (\ref{l3:eq1a}). For model (a), the flux is computed via the $E_{phi}$ controlled voltage source as a product of this meminductance and the current $I$ flowing through the meminductive port. The port voltage is then evaluated as time derivative of this flux (see the controlled voltage source $E_L$). For model (b), the flux is computed via integration of the port voltage, and the port current is derived as a ratio of the flux and the meminductance, thus obeying the differentiation.

In PSpice and LTSpice, both models work well. However, HSPICE operates only with the model (a) whereas convergence problems are reported for model (b). They can be overcome after running HSPICE RF instead of HSPICE.
Appendix \ref{app_L3} summarizes SPICE codes for more reliable model in Fig. \ref{figL3a}(a). A demonstration of PSpice outputs is shown in Fig. \ref{figL3b}. It can be observed that the low level of the meminductance is not $L_\text{low}$ but it is preserved to the initial value $L_{init}$ (see Fig. \ref{figL3b} and SPICE code in Appendix \ref{app_L3}). The boundary value is switched to $L_\text{low}$ after increasing the magnitude of the exciting current above a proper value.

%Possible parameters: $L_{low}=1\mu$H, $L_{high}=100\mu$H, $I_t=10\mu$A, $\beta=10^7\mu$H/(V s). These parameters assume $\mu$s switching times \textbf{Are these reasonable values for $L$ in microstructures?}.

\section{Setting the analysis parameters and SPICE options} \label{secC}

In this Section we discuss several rules for solving accuracy and convergence problems in SPICE via tweaking analysis parameters and global settings. The common rules are described in a number of references including a couple of excellent books \cite{kundert1995designer,kielkowski1998inside}. Some of the rules discussed below are focused on the specifics of memelement simulation within the transient analysis, which is most frequently used for this type of components.

Incorrect modeling is a common source of the problems burdening the transient analysis. The rules of building-up correct models of memsystems have been described in Section \ref{sec2B}, thus they will not be dealt with below.
The problems appearing within the analysis can be of the following two types.
Convergence problems: SPICE does not find the solution (fatal problems indicating by error message).
Accuracy problems: The solution is found but it is modified by errors (problems which can be hidden particularly if we have no idea of the correct result).
Since the attempts at increasing the accuracy attracts the convergence problems, the transient analysis of systems requiring extremely high accuracy can be considered as art of compromise. Ideal memelements or memristive systems with threshold are typical representatives of the above systems (see models R.1, R.2, C.1, C.4, L.1 and L.3 in Sections \ref{sec3}-\ref{sec5}).
The above convergence and accuracy problems, if they appear, must be handled in the sequence as they are mentioned. If the circuit does not converge, one cannot deal with the accuracy of the solution.

Note that the SPICE command for the transient analysis can be in one of two basic forms:
\lstset{xleftmargin=10pt}
\begin{lstlisting}
.TRAN Tprint Tstop [skipbp]             (*)
   or
.TRAN Tprint Tstop Tstart Hmax [skipbp] (**)
\end{lstlisting}
\lstset{xleftmargin=20pt}
with an optional flag skipbp or uic. In addition to the commands (*), (**), the algorithms of the analysis can be affected by the attributes defined by the .OPTIONS command, especially the error and other iteration criteria.
The transient analysis has two stages, the DC bias point calculation and the timepoint sweep analysis. The analysis result depends on the behavior of the numerical algorithms acting in both stages. The first stage can be skipped via the skipbp flag although it is not generally recommended \cite{kundert1995designer}.

\vspace{0.3cm}

\noindent {\bf Convergence aids for DC bias point computation}

If SPICE fails to converge to a DC bias point, it aborts the Newton-Raphson (NR) iteration and prints the error message "No convergence in DC operating point". Note that most simulation examples from Sections \ref{sec3} to \ref{sec5} with SPICE codes from the Appendices, work without any convergence problems, since their models were built up according to rules from Section \ref{sec2B}.
The main difficulties are related to implementations of C.4. and L.3 in HSPICE. Moreover, the convergence strongly depends on application deteriorating in circuits leading to large sets of equations. The suggested sequence of the actions is summarized below.

\begin{enumerate}
\item Raise ITL1, i.e. the upper iteration limit of the Newton-Raphson (NR) method from its default value 150 to 500 or more via the command \newline
.OPTION ITL1=500
\item Via the .NODESET command, set the qualified estimation of DC values of as many nodal voltages as possible.
\item Call the Source Stepping algorithm via the command \newline
.OPTION ITL6=500
\item Increase the GMIN parameter above its default value $10^{-12}\Omega^{-1}$, for example \newline
.OPTION GMIN=1E-10 \newline
GMIN is estimated as reciprocal value of the smallest parasitic resistance which could be placed across any two nodes without influencing the model behavior \cite{kielkowski1998inside}.
\item Consider if the relative error of voltages and currents can be higher than the default value 0.1\%. If yes, then raise RELTOL: \newline
.OPTION RELTOL=0.01
\item Determine the magnitude of the smallest voltage of interest, e.g. 1uV, and compute the absolute voltage error VNTOL=RELTOL*1u =1E-8. Then redefine VNTOL from its default value 1uV: \newline
.OPTION VNTOL=1E-8 \newline
If you cannot estimate the smallest voltage, then use the rule that VNTOL should be by 6 to 9 orders smaller than the largest voltage in the circuit \cite{kundert1995designer}.
\item Determine the magnitude of the smallest current of interest, e.g. 1uA, and compute the absolute current error ABSTOL=RELTOL*1u =1E-8. Then redefine ABSTOL from its default value 1pA: \newline
.OPTION ABSTOL=1E-8 \newline
If you cannot estimate the smallest current, then use the rule that ABSTOL should be by 6 to 9 orders smaller than the largest current in the circuit \cite{kundert1995designer}.
\item If the above hints do not help, remove the skipbp from the .TRAN command. If concrete nodal voltages can be estimated, define them via .IC command
\end{enumerate}

Note that steps 4-7 solve the convergence problems at the expense of the accuracy.
In addition to SPICE standard, HSPICE offers additional convergence aids, particularly "Modified Source-Stepping Algorithm" (MSSA), "Gmin Ramping" (GMR), and "Pseudo-Transient Analysis" (PTA). MSSA, which can be enabled via the .OPTION CONVERGENCE=3, can be used instead of step No. 3). GMR can replace the step No. 4). It can be initiated as .OPTION GRAMP=X where X is for example 6. PTA is an efficient generalization of the step No. 3). It can be activated via the command .OPTION CONVERGENCE=1. Details are available in \cite{HSPICE08a}.

If PSpice fails to converge within ITL1 limit, the Source Stepping Algorithm (SSA) is switched on automatically, without a possibility of controlling this process by the user. If SSA also fails to converge, the Gmin Stepping (ramping) can be initiated via the command .OPTION STEPGMIN. This method is then applied first, and if it will not converge, PSpice comes to SSA algorithm.
If LTspice does not converge, it tries the algorithms of adaptive GMR, adaptive SSA, and PTA in successive steps. The user can deactivate individual algorithms from the queue via the corresponding flags \cite{Micro10a}.

\vspace{0.3cm}

\noindent {\bf Convergence aids for timepoint sweep analysis}

The DC bias solution is a starting point of the transient analysis which computes the solution at timepoints via numerical integration of circuit equations. The methods of the numerical integration used in SPICE are Backward Euler (BE), Trapezoidal (TRAP), and Gear (GEAR)). LTspice offers TRAP, GEAR2 (i.e. second-order GEAR), a special modification of TRAP, and BE (it is initiated by undocumented command .OPTIONS MAXORD=1). PSpice uses only TRAP combined with BE. HSPICE provides TRAP and GEAR of orders 1 to 6, with GEAR1 being the BE method.

The circuit solution at each timepoint is found via the NR iteration. The timepoints are not evenly spaced on the time axis but their density is controlled via the TimeStep Control (TSC) algorithm depending on how fast the circuit voltages and currents are moving. The SPICE standard defines two methods of timestep control, Iteration Count (IC) and Local Truncation Error (LTE). In addition, HSPICE offers the third method called DVDT Dynamic Timestep \cite{HSPICE08a}. PSpice and LTspice use only LTE method.

The convergence problems appear as a consequence of the simultaneous action of NR and TSC algorithms. They are accompanied by "Internal Timestep Too Small" or "No Convergence During Transient Analysis" error messages, indicating that the solution was not found even though the timestep reached its minimum allowable limit.
In the first step, it should be checked if the improper model of the circuit is not the key source of convergence problems (see Section \ref{sec2B}).
The other recommended steps are summarized below.

\begin{enumerate}
\item Raise ITL4, i.e. the upper iteration limit at each timepoint, to 50 or more via the command \newline
.OPTION ITL4=50
\item Select GEAR2 integration method (not for PSpice).
\item Loosen error criteria of NR algorithm according to steps 5-7 from the Convergence aids for DC bias point computation.
\item Increase TRTOL tentatively above its default value (7 for HSPICE, 1 for LTspice).
\end{enumerate}

\vspace{0.3cm}

\noindent {\bf Accuracy aids for transient analysis of memsystems}

After resolving prospective convergence problems, the options of NR and TSC algorithms can be tweaked to maximize the accuracy. Note that there are two fundamental limits of the accuracy increase:
\begin{itemize}
\item	numerical limits in the representation of voltages, currents, and system variable TIME as well as numerical noise which can be amplified or accumulated by the circuit model (see Section \ref{sec2B}, Item 3),
\item	increase of the accuracy promotes the convergence problems.
\end{itemize}
SPICE provides the following options for increasing the accuracy of the transient analysis:
Selection of the integration method (not in PSpice), selection of the type of TSC algorithm (not in PSpice and LTspice) and its parameters, selection of the parameters of the NR algorithm and .TRAN command options.
In addition to the above tools, it is important to find a suitable guideline for checking the correctness and the accuracy of the analysis of concrete systems. Demonstrations of such guidelines, which start from the fundaments of the analyzed memelements, in particular of ideal memristor (R.1) and memcapacitor (C.1) or bipolar memristive system (R.2) are given in Sections \ref{sec3} and \ref{sec4}.

In virtue of the experience in SPICE simulation of assorted types of memristive, memcapacitive and meminductive systems, the key factors influencing the accuracy of the transient analysis are identified and summarized in the following steps.

\begin{enumerate}
\item Tighten RELTOL below its default value 0.001. Set the other error criteria according to steps 6 and 7 from the Convergence aids for DC bias point computation.
\item Analyze if the default value of Gmin=10$^{-12}$ does not affect the accuracy. If possible, set Gmin=0 \cite{kundert1995designer}.
\item LTspice, HSPICE: Select GEAR2 as integration method.
\item The parasitic ringing generated by TRAP method \cite{kielkowski1998inside} can be solved either by switching to GEAR2 or via step No. 7.
\item The parasitic overshoot generated by GEAR2 method \cite{kielkowski1998inside} can be solved via step No. 7.
\item Accumulated errors (divergence from the correct solution during long transient run) \cite{kielkowski1998inside} can be solved via step No. 7.
\item Tighten maximum timestep Tmax (via parameter Hmax or Tprint, see details below).
\item Tweaking the options of LTE algorithm of dynamic timestep control \cite{kundert1995designer} (via TRTOL or RELTOL, see details below).
\item HSPICE: Select the algorithm of dynamic timestep control and its parameters \cite{HSPICE08a} (see details below).
\end{enumerate}

The recommendation 3) is based on the practice that GEAR2 method is suitable for the analysis of memelements of various natures. Since the models of memory systems contain ideal integrators, GEAR2 is a good choice owing to its stable behavior when evaluating integrals of circuit quantities within many repeating periods. In addition, GEAR2 provides good results for stiff systems where the signals move too fast with respect to the actual timestep size. Typical cases are voltage-controlled memcapacitive systems or current-controlled meminductive systems where the port quantities, namely the capacitor current and inductor voltage, are computed via numerical differentiation of controlling voltage and current. Examples are given in Sections \ref{sec4} and \ref{sec5} under the codes C.1 and L.1. Though the BE method is the best for stiff systems, we should avoid it because it accumulates errors when analyzing integration blocks. PSpice does not provide Gear integration. Fortunately, PSpice combines TRAP with BE, this way eliminating trapezoidal oscillations and its negative effects. For memelements with hard-switching effects and other systems which exhibit fast signal transitions, GEAR2 behaves well with regard to the accumulated errors. Bipolar memristive system R.2 from Section \ref{sec3} is a typical representative of such systems exhibiting the switching phenomena.

It turns out from the above that GEAR2 can be optimal choice for memelements. The possible imperfections can be suppressed by decreasing the maximum step size (see below). On the other hand, Gear method may not provide the best results, and the standard trapezoidal algorithm can solve the task in some cases (see memristive systems R.3 and R.4, memcapacitive systems C.2 and C.3, and meminductive system L.2 in Sections \ref{sec3} to \ref{sec5}). If we can select among the methods, then it is useful to try out the model behavior with all the methods and to face the results with the expected waveforms.

After selecting the integration algorithm, increasing the accuracy and eliminating the inherent parasitic behavior of the method can be accomplished via tightening the timestep. An indirect method of tightening the timestep is decreasing the maximum timestep Tmax (see Item 7 in the above steps). In PSpice and LTspice, which utilize the LTE method of dynamic timestep control, Tmax is set as Tmax=MIN(Tstop/50, Hmax). The extended syntax (**) of the .TRAN command should be used with Hmax small enough (e.g. Tstop/1000). In HSPICE, the stepsize control is rather complicated. Tmax can be set via Tprint which appears in the simple syntax (*) of .TRAN command. Note that it depends also on other flags such as RMAX. HSPICE also provides direct Tmax control via the flag DELMAX. See \cite{HSPICE08a} for details.

If the LTE method of dynamic step control is used, then the step size can be tightened directly via error criteria, particularly TRTOL and RELTOL (see Item 8 in the above steps). The size of the actual step is proportional to the root of the product of TRTOL and RELTOL \cite{kielkowski1998inside}. Tightening RELTOL (see Item 1 in the above steps) improves the precision of both NR and integration algorithms. Tightening TRTOL refines only the integration method without influencing NR algorithm. TRTOL default value is 7 for PSpice and 1 for LTspice, thus LTspice should produce ca 2.6 times (root of 7) smaller timestep than PSpice. Even if lowering TRTOL much below its default value is not generally recommended \cite{kundert1995designer}, this method can significantly improve the accuracy. Section III demonstrates one example R.4 (Insulator-to-metal transition memristive system) where TRTOL=0.1 provides the regime of enhanced precision for LTspice. Similar effect can provide the option RELTOL=1u for PSpice. Refer to \cite{kundert1995designer} for more details about the accuracy issue related to LTE method.

HSPICE offers inexhaustible options of improving the accuracy of the analysis of memelements. It enables combination of various integration methods, algorithms of dynamic timestep control, and error criteria. In this sense, it goes far beyond the SPICE standard. The so-called RUNLVL algorithms with 6 discrete levels (1-fastest, 6-most accurate) can be used for simplifying the optimization of transient analysis. These algorithms use LTE method for dynamic timestep control. The command .OPTION RUNLVL=6 is used in the source code for the simulation of bipolar memristive system R.2 from Section \ref{sec3} to provide high precision of computing time instants of switching the memristance states.

HSPICE provides excellent performance for complex semiconductor devices but it sometimes fails when analyzing behavioral models based on formulae and controlled sources. Two examples are given in Sections \ref{sec4} and \ref{sec5} (bipolar memcapacitive and meminductive systems with threshold). The "Golden Reference for Options" is recommended for finding the acceptable trade-off between HSPICE accuracy and transient analysis simulation performance \cite{HSPICE08b}:
\begin{lstlisting}
.OPTION RUNLVL=6 ACCURATE KCLTEST
+ DELMAX=<something_small>
\end{lstlisting}
Via this option, DELMAX can be decreased tentatively in order to acquire as accurate results as HSPICE allows (see the HSPICE codes of threshold devices C.4 and L.3 in Appendices \ref{app_C4} and \ref{app_L3}). The flag KCLTEST activates Kirchhoff's Current Law for every circuit node via tightening the error criteria. Note that it was used for increasing the accuracy of the simulation of meminductive threshold device L.3 (see the Appendix \ref{app_L3}). The flag ACCURATE sets additional HSPICE options to stricter tolerances. See \cite{HSPICE08b} for more details.

Although the LTE algorithm is allowed to be more precise than IC method of timestep control \cite{kundert1995designer,kielkowski1998inside}, it generates inaccurate results for some types of circuits containing memelements. It relates to the circuits employing ideal memristors, memcapacitors, and meminductors (see examples R.1, C.1, and L.1 in Sections \ref{sec3} to \ref{sec5}). HSPICE controls the timestep by means of a complicated mix of DVDT, IC and LTE algorithms. The type of the method is set by the flag LVLTIM, but the RUNLVL algorithm must be disabled first via the command .OPTION RUNLVL=0. For circuits containing ideal memelements, the DVDT algorithm in combination with IC algorithm is the choice which provides most accurate analysis. It can be set via the command .OPTION LVLTIM=1 (see the HSPICE source codes for circuits R.1, C.1, and L.1).

\section{Conclusions}
In summary, we have presented a coherent approach to reliably simulate memristive, memcapacitive, and
meminductive systems in the SPICE environment. Apart from general considerations on the ``best practices''
to carry out the simulations for these particular devices, we have provided a lot of examples
for all three classes of memelements. For the benefit of the reader, we have also provided in the
Appendices many codes of these models written in the most popular SPICE versions (PSpice, LTspice, HSPICE) that can be simply ``cut and paste'' in the appropriate environment for immediate
test and execution. Our goal would be accomplished if we could help researchers build from our own experience, avoid common pitfalls in the simulation of these new devices, and venture into their own simulations.

\begin{widetext}

\appendices

\section{SPICE codes for model R.1} \label{app_R1}

\noindent{\bf PSpice and LTspice code}

\begin{lstlisting}
**** Ideal memristor model R1 ****
*D. Biolek, M. Di Ventra, Y. V. Pershin*
*Reliable SPICE Simulations of Memristors, Memcapacitors and Meminductors, 2013*
*Code for PSpice and LTspice; tested with Cadence PSpice v. 16.3 and LTspice v. 4*
**********************************************************************
.subckt memristorR1 plus minus params: Ron=100 Roff=10k Rini=5k
.param uv=10f D=10n k={uv*Ron/D**2} a={(Rini-Ron)/(Roff-Rini)}
*model of memristive port
Roff plus aux {Roff}
Eres aux minus value={(Ron-Roff)/(1+a*exp(-4*k*V(q)))*I(Eres)}
*end of the model of memristive port
*integrator model
Gx 0 Q value={i(Eres)}
Cint Q 0 1
Raux Q 0 100meg
*end of integrator model
*alternative integrator model; SDT function for PSPICE can be replaced by IDT for LTspice
*Eq Q 0 value={SDT(I(Eres))}
.ends memristorR1

*.options method=gear  ;use only for LTSpice
Vin in 0 sin 0 1 1
Xmem in 0 memristorR1
.tran 0 10 0 1m
.probe
.end
\end{lstlisting}

\noindent{\bf HSPICE code}

\begin{lstlisting}
**** Ideal memristor model R1 ****
*D. Biolek, M. Di Ventra, Y. V. Pershin*
*Reliable SPICE Simulations of Memristors, Memcapacitors and Meminductors, 2013*
*Code for HSPICE; tested with HSPICE Version A-2008.03*
**********************************************************************
.subckt memristorR1 plus minus Ron=100 Roff=10k Rini=5k
.param uv=10f D=10n k='uv*Ron/D**2' a='(Rini-Ron)/(Roff-Rini)'
*model of memristive port
Roff plus aux 'Roff'
Eres aux minus vol='(Ron-Roff)/(1+a*exp(-4*k*V(q)))*I(Eres)'
*end of the model of memristive port
*integrator model
Gx 0 Q cur='i(Eres)'
Cint Q 0 1
Raux Q 0 100meg
*end of integrator model
.ends memristorR1

.options post runlvl=0 lvltim=1 method=gear
Vin in 0 sin(0,1,1)
Xmem in 0 memristorR1
.tran 0.1m 10
.probe v(x*.*) i(x*.*)
.end
\end{lstlisting}

\section{SPICE codes for model R.2} \label{app_R2}

\noindent{\bf PSpice and LTspice code}

\begin{lstlisting}
**** Bipolar memristive system with threshold R2 ****
*D. Biolek, M. Di Ventra, Y. V. Pershin*
*Reliable SPICE Simulations of Memristors, Memcapacitors and Meminductors, 2013*
*Code for PSpice and LTspice; tested with Cadence PSpice v. 16.3 and LTspice v. 4*
**********************************************************************
.subckt memR_TH plus minus PARAMS:
+ Ron=1K Roff=10K Rinit=5K beta=1E13 Vt=4.6
*model of memristive port
Gpm plus minus value={V(plus,minus)/V(x)}
*end of the model of memristive port
*integrator model
Gx 0 x value={fs(V(plus,minus),b1)*ws(v(x),V(plus,minus),b1,b2)*1p}
Raux x 0 1T
Cx x 0 1p IC={Rinit}
*end of integrator model
*smoothed functions
.param b1=10u b2=10u
.func stps(x,b)={1/(1+exp(-x/b))}
.func abss(x,b)={x*(stps(x,b)-stps(-x,b))}
.func fs(v,b)={beta*(v-0.5*(abss(v+Vt,b)-abss(v-Vt,b)))}
.func ws(x,v,b1,b2)={stps(v,b1)*stps(1-x/Roff,b2)+stps(-v,b1)*stps(x/Ron-1,b2)}
*end of smoothed functions
.ends memR_TH

.options reltol=1u
*.options method=gear ;use only for LTspice
Vsin 1 0 sin 0 5 50meg
Xmem 1 0 memR_TH
.tran 0 0.1u 0 0.1n
.probe
.end
\end{lstlisting}

\noindent{\bf HSPICE code}

\begin{lstlisting}
**** Bipolar memristive system with threshold R2 ****
*D. Biolek, M. Di Ventra, Y. V. Pershin*
*Reliable SPICE Simulations of Memristors, Memcapacitors and Meminductors, 2013*
*Code for HSPICE; tested with HSPICE Version A-2008.03*
**********************************************************************
.subckt memR_TH plus minus
+ Ron=1K Roff=10K Rinit=5K beta=1E13 Vt=4.6
*model of memristive port
Gpm pl mn cur='V(plus,minus)/V(x)'
*end of the model of memristive port
*integrator model
Gx 0 x cur='fs(V(plus,minus),b1)*ws(v(x),V(plus,minus),b1,b2)*1p'
Raux x 0 1T
Cx x 0 1p
.IC v(x)='Rinit'
*end of integrator model
*smoothed functions
.param b1=10u b2=10u
.param stps(x,b)='1/(1+exp(-x/b))'
.param abss(x,b)='x*(stps(x,b)-stps(-x,b))'
.param fs(v,b)='beta*(v-0.5*(abss(v+Vt,b)-abss(v-Vt,b)))'
.param ws(x,v,b1,b2)='stps(v,b1)*stps(Roff-x,b2)+stps(-v,b1)*stps(x-Ron,b2)'
*end of smoothed functions
.ends memR_TH

.option post runlvl=6 method=gear
Vsin 1 0 sin(0,5,50meg)
Xmem 1 0 memR_TH
.tran 0.1n 0.1u
.probe v(x*.*) i(x*.*)
.end
\end{lstlisting}

\section{SPICE code for model R.3} \label{app_R3}

\noindent{\bf PSpice and LTspice code}

\begin{lstlisting}
**** Phase change memristive system R3 ****
*D. Biolek, M. Di Ventra, Y. V. Pershin*
*Reliable SPICE Simulations of Memristors, Memcapacitors and Meminductors, 2013*
*Code for PSpice and LTspice; tested with Cadence PSpice v. 16.3 and LTspice v. 4*
**********************************************************************
.subckt PCM plus minus PARAMS:
+ Ron=10K Roff=1meg Rini=100k alpha=20meg beta=100meg
+ Tr=20 Tx=200 Tm=600 Tini=20 Ch=2e-15 d=5u
+ Vtr=1.8 V0=50m Cxini=0
*resistive port modeling
Ron plus aux {Ron}
Eres aux minus value={(Roff-Ron)*(1-V(Cx))/(1+exp((V(plus,minus)-Vtr)/V0))*I(Eres)}
*end of resistive port modeling
*temperature computation
GT 0 T value={V(plus,minus)*I(Eres)+d*(Tr-V(T))}
RauxT T 0 100meg
CintT T 0 {Ch} IC={Tini}
*end of temperature computation
*Cx computation
GCx 0 Cx value=
+ {alpha*(1-V(Cx))*stps(V(T)/Tx-1)*stps(1-V(T)/Tm)-beta*V(Cx)*stps(V(T)/Tm-1)}
RauxCx Cx 0 100meg
CintCx Cx 0 1 IC={Cxini}
*end of Cx computation
*smoothed step function
.param b=1m
.func stps(x)={1/(1+exp(-x/b))}
*end of smoothed step function
.ends PCM

V 1 0 PWL
+ 0 4 300n 4 301n 0 400n 0 401n 6 500n 6 501n 0
Xmem 1 0 PCM
.tran 0 600n
.probe
.end
\end{lstlisting}

\noindent{\bf HSPICE code}

\begin{lstlisting}
**** Phase change memristive system R3 ****
*D. Biolek, M. Di Ventra, Y. V. Pershin*
*Reliable SPICE Simulations of Memristors, Memcapacitors and Meminductors, 2013*
*Code for HSPICE; tested with HSPICE Version A-2008.03*
**********************************************************************
.subckt PCM plus minus
+ Ron=10K Roff=1meg Rini=100k alpha=20meg beta=100meg
+ Tr=20 Tx=200 Tm=600 Tini=20 Ch=2e-15 d=5u
+ Vtr=1.8 V0=50m Cxini=0
*resistive port modeling
Ron plus aux 'Ron'
Er aux minus vol='(Roff-Ron)*(1-V(Cx))/(1+exp((V(plus,minus)-Vtr)/V0))*I(Er)'
*end of resistive port modeling
*temperature computation
GT 0 T cur='V(plus,minus)*I(Er)+d*(Tr-V(T))'
RauxT T 0 100meg
CintT T 0 'Ch'
.IC v(T)='Tini'
*end of temperature computation
*Cx computation
GCx 0 Cx cur=
+ 'alpha*(1-V(Cx))*stps(V(T)/Tx-1)*stps(1-V(T)/Tm)-beta*V(Cx)*stps(V(T)/Tm-1)'
RauxCx Cx 0 100meg
CintCx Cx 0 1
.IC v(Cx)='Cxini'
*end of Cx computation
*smoothed step function
.param b=1m
.param stps(x)='1/(1+exp(-x/b))'
*end of smoothed step function
.ends PCM

.option post
V 1 0 PWL
+ 0 4 300n 4 301n 0 400n 0 401n 6 500n 6 501n  0
Xmem 1 0 PCM
.tran 6n 600n
.probe v(x*.*) i(x*.*)
.end
\end{lstlisting}

%{\bf Parameters:}
%$R_\text{on}=10^4 \Omega$, $R_\text{off}=10^6\Omega$, $T_\text{r}=20\,^{\circ}\mathrm{C}$, $T_\text{m}=600\,^{\circ}\mathrm{C}$,
%$T_\text{x}=200\,^{\circ}\mathrm{C}$, $\alpha=2\cdot 10^{7}$ s$^{-1}$,  $\beta=10^{8}$ s$^{-1}$, $C_\text{h}=2 \cdot 10^{-15}$J/K, $\delta=5 \cdot 10^{-6}$W/K, $V_\text{t}=1.8$V, %$V_0=0.05$V. The selected set of parameters is based on Refs. \cite{Agarwal07a,ventrice2007phase,dao2011compact}.

\section{SPICE code for model R.4} \label{app_R4}

\noindent{\bf PSpice and LTspice code}
%\textbf{Definition of pi was added}

\begin{lstlisting}
**** Insulator-to-metal transition memristive system R4 ****
*D. Biolek, M. Di Ventra, Y. V. Pershin*
*Reliable SPICE Simulations of Memristors, Memcapacitors and Meminductors, 2013*
*Code for PSpice and LTspice; tested with Cadence PSpice v. 16.3 and LTspice v. 4*
**********************************************************************
.subckt IMTM plus minus PARAMS: uini=1u
.param deltaT=784 rch=30n L=20n Rhoins=7m Rhomet=100u
+ deltaHtr=1.6e8 k=1.5 cp=2.6meg
.func Gammath(u)={-2*pi*L*k/log(u)}
.func dHdu(u)={pi*L*rch**2*(cp*deltaT*uExpr(u)+2*deltaHtr*u)}
.func uExpr(u)={(1-u**2+2*u**2*log(u))/(2*u*log(u))**2}
*resistive port modeling
Vsense plus sense 0
Rfix sense minus {Rhoins*L/(pi*rch**2)}
Gvar sense minus value={V(plus,minus)*v(uL)**2*pi*rch**2/L*(1/Rhomet-1/Rhoins)}
*end of resistive port modeling
*u computation
Gu 0 u value={1p/dHdu(v(uL))*(v(plus,minus)*I(Vsense)-Gammath(v(uL))*deltaT)}
Raux u 0 10G
Cu u 0 1p IC={uini}
*end of u computation
*u limits
EuL uL 0 value={LIMIT(v(u),1u,0.99999)}
*end of u limits
.ends IMTM

*modeling Pearson-Anson relaxation oscillator
*.options trtol=0.1 method=gear ; use only in LTSpice
.options reltol=1u
Vdc 1 0 1.8
RL 1 2 4.2k
Re 2 3 2.7k
Rscope 4 0 50
Cp 2 0 23p
XIMTM 3 4 IMTM
.tran 0 10u 8u 1n
.probe
.end
\end{lstlisting}

\noindent{\bf HSPICE code}

\begin{lstlisting}
**** Insulator-to-metal transition memristive system R4 ****
*D. Biolek, M. Di Ventra, Y. V. Pershin*
*Reliable SPICE Simulations of Memristors, Memcapacitors and Meminductors, 2013*
*Code for HSPICE; tested with HSPICE Version A-2008.03*
**********************************************************************
.subckt IMTM plus minus uini=1u
.param deltaT=784 rch=30n L=20n Rhoins=7m Rhomet=100u
+ deltaHtr=1.6e8 k=1.5 cp=2.6meg
.param pi=3.1415926536
.param Gammath(u)='-2*pi*L*k/log(u)'
.param dHdu(u)='pi*L*rch**2*(cp*deltaT*uExpr(u)+2*deltaHtr*u)'
.param uExpr(u)='(1-u**2+2*u**2*log(u))/(2*u*log(u))**2'
*resistive port modeling
Vsense plus sense 0
Rfix sense minus 'Rhoins*L/(pi*rch**2)'
Gvar sense minus cur='V(plus,minus)*v(uL)**2*pi*rch**2/L*(1/Rhomet-1/Rhoins)'
*end of resistive port modeling
*u computation
Gu 0 u cur='1p/dHdu(v(uL))*(v(plus,minus)*I(Vsense)-Gammath(v(uL))*deltaT)'
Raux u 0 10G
Cu u 0 1p
.IC v(u)='uini'
*end of u computation
*u limits
EuL uL 0 vol='min(max(v(u),1u),0.99999)'
*end of u limits
.ends IMTM

*modeling Pearson-Anson relaxation oscillator
.option post runlvl=6
Vdc 1 0 1.8
RL 1 2 4.2k
Re 2 3 2.7k
Rscope 4 0 50
Cp 2 0 23p
XIMTM 3 4 IMTM
.tran 1n 10u 8u 1n
.probe v(x*.*) i(x*.*)
.end
\end{lstlisting}

%{\bf Parameters:}
%See Table I in Ref. \cite{pickett2012sub} for a detailed list of parameters.

\section{SPICE code for model C.1} \label{app_C1}

\noindent{\bf PSpice and LTspice code}

\begin{lstlisting}
**** Ideal memcapacitor C1 ****
*D. Biolek, M. Di Ventra, Y. V. Pershin*
*Reliable SPICE Simulations of Memristors, Memcapacitors and Meminductors, 2013*
*Code for PSpice and LTspice; tested with Cadence PSpice v. 16.3 and LTspice v. 4*
**********************************************************************
.subckt memcapacitor plus minus params: Clow=1p Chigh=100p Cini=2p k=100
.param a={(Chigh-Cini)/(Cini-Clow)}
*model of memcapacitive port
.func C(phi)={Clow+(Chigh-Clow)/(a*exp(-4*k*phi)+1)}
EQ Q 0 value={C(V(phi))*V(plus,minus)}
Gcap plus minus value={ddt(V(Q))}
*for OrCAD PSpice 16, the above line can be replaced by Gcap plus minus Q={V(Q)}
*end of the model of memcapacitive port
*integrator model
Gv 0 phi value={v(plus,minus)}
Cint phi 0 1
Raux phi 0 100meg
*end of integrator model
.ends memcapacitor

Vin 1 0 sin 0 1 10
XMC 1 0 memcapacitor
.tran 0 0.2 0 1m
.probe
.end
\end{lstlisting}

\noindent{\bf HSPICE code}

\begin{lstlisting}
**** Ideal memcapacitor C1 ****
*D. Biolek, M. Di Ventra, Y. V. Pershin*
*Reliable SPICE Simulations of Memristors, Memcapacitors and Meminductors, 2013*
*Code for HSPICE; tested with HSPICE Version A-2008.03*
**********************************************************************
.subckt memcapacitor plus minus Clow=1p Chigh=100p Cini=2p k=100
.param a='(Chigh-Cini)/(Cini-Clow)'
*model of memcapacitive port
.param C(phi)='Clow+(Chigh-Clow)/(a*exp(-4*k*phi)+1)'
EQ Q 0 vol='C(V(phi))*V(plus,minus)'
CQ plus minus C='C(V(phi))' CTYPE=1
*end of the model of memcapacitive port
*integrator model
Gv 0 phi cur='v(plus,minus)'
Cint phi 0 1
Raux phi 0 100meg
*end of integrator model
.ends memcapacitor

.option runlvl=0 lvltim=1 method=gear
Vin 1 0 sin(0,1,10)
XMC 1 0 memcapacitor
.tran 1m 0.2
.probe v(x*.*) i(x*.*)
.end
\end{lstlisting}

\section{SPICE code for model C.2} \label{app_C2}

\noindent{\bf PSpice and LTspice code}

\begin{lstlisting}
**** Multilayer memcapacitive system C2 ****
*D. Biolek, M. Di Ventra, Y. V. Pershin*
*Reliable SPICE Simulations of Memristors, Memcapacitors and Meminductors, 2013*
*Code for PSpice and LTspice; tested with Cadence PSpice v. 16.3 and LTspice v. 4*
**********************************************************************
.subckt MLMCS plus minus params: d=100n del=66.6n Su=100u er=5 Uev=0.33
.param e0=8.854p m=9.109e-31 e=1.602e-19 h=6.626e-34
.param C0={e0*er*Su/d} C1={C0/(1-del/d)} C2={C0*d/del}
*Use this below line for LTSpice
*.param a={Su*e**2/(4*pi*h*Uev*del**2)} b={4*pi*del*sqrt(m*e)*pwr(Uev,1.5)/h} loga={log(a)}
*Use this below line for OrCAD PSpice
.param a=2.10572e5 b=91.4682096 loga={log(a)}
.func I12(V1)={V1*abs(V1)*exp(LIMIT(loga-b/MAX(abs(V1),1n),-20,20))}
*model of memcapacitive port
C1 plus c {C1}
C2 c minus {C2}
GQ c minus value={I12(V(c,minus))}
Rshunt c 0 100meg
*end of the model of memcapacitive port
.ends MLMCS

Vin 1 0 sin 0 7.5 100
Rin 1 2 1
XMC 2 0 MLMCS
EQ Q 0 value={-sdt(I(Vin))}
.tran 0 50m 20m 50u skipbp
.probe
.end
\end{lstlisting}

\noindent{\bf HSPICE code}

\begin{lstlisting}
**** Multilayer memcapacitive system C2 ****
*D. Biolek, M. Di Ventra, Y. V. Pershin*
*Reliable SPICE Simulations of Memristors, Memcapacitors and Meminductors, 2013*
*Code for HSPICE; tested with HSPICE Version A-2008.03*
**********************************************************************
.subckt MLMCS plus minus d=100n del=66.6n Su=100u er=5 Uev=0.33
.param pi=3.1415926536 e0=8.854p m=9.109e-31 e=1.602e-19 h=6.626e-34
.param C0='e0*er*Su/d' C1='C0/(1-del/d)' C2='C0*d/del'
.param a=2.10572e5 b=91.4682096 loga='log(a)'
.param I12(V1)='V1*abs(V1)*exp(min(max(loga-b/MAX(abs(V1),1n),-20),20))'
*model of memcapacitive port
C1 plus c 'C1'
C2 c minus 'C2'
GQ c minus cur='I12(V(c,minus))'
Rshunt c 0 100meg
*end of the model of memcapacitive port
.ends MLMCS

.option post
Vin 1 0 sin(0,7.5,100)
Rin 1 2 1
XMC 2 0 MLMCS
*charge computation
Gqq qq 0 vol='I(Vin)'
Cqq qq 0 1
Rqq qq 0 100meg
*end of charge computation
.tran 50u 50m 20m 50u
.probe v(x*.*) i(x*.*)
.end
\end{lstlisting}

\section{SPICE code for model C.3} \label{app_C3}

\noindent{\bf PSpice and LTspice code}
%\textbf{Definition of pi was added}
\begin{lstlisting}
**** Bistable membrane memcapacitive system C3 ****
*D. Biolek, M. Di Ventra, Y. V. Pershin*
*Reliable SPICE Simulations of Memristors, Memcapacitors and Meminductors, 2013*
*Code for PSpice and LTspice; tested with Cadence PSpice v. 16.3 and LTspice v. 4*
**********************************************************************
.subckt BEMS plus minus params: y0=0.2 yd0=0
.param gamma=0.7 b=1 C0=10p
*model of memcapacitive port
C0 plus c {C0}
Ec c minus value={V(Q)*V(y)/C0}
*end of the model of memcapacitive port
*Q computation
EQ Q 0 value={C0*V(plus,c)}
*end of Q computation
*y computation
Gy 0 y value={v(yd)}
Cy y 0 1 IC={y0}
Ry y 0 100meg
*end of y computation
*yd computation
Gyd 0 yd value={-(4*pi**2*v(y)*((V(y)/y0)**2-1)+gamma*v(yd)+(b*V(plus,minus)/(1+v(y)))**2)}
Cyd yd 0 1 IC={yd0}
Ryd yd 0 100meg
*end of yd computation
.ends BEMS

Vin 1 0 sin 0 2.8 0.658
Rin 1 2 1
XMC 2 0 BEMS
.tran 0 20 16 4m
.probe
.end
\end{lstlisting}

\noindent{\bf HSPICE code}

\begin{lstlisting}
**** Bistable membrane memcapacitive system C3 ****
*D. Biolek, M. Di Ventra, Y. V. Pershin*
*Reliable SPICE Simulations of Memristors, Memcapacitors and Meminductors, 2013*
*Code for HSPICE; tested with HSPICE Version A-2008.03*
**********************************************************************
.subckt BEMS plus minus y0=0.2 yd0=0
.param pi=3.1415926536 gamma=0.7 b=1 C0=10p
*model of memcapacitive port
C0 plus c 'C0'
CQ c minus C='C0/V(y)' CTYPE=1
*Ec c minus vol='V(Q)*V(y)/C0'
*end of the model of memcapacitive port
*Q computation
*GQ 0 Q cur='I(EC)'
*CQ Q 0 1
*RQ Q 0 100meg
EQ Q 0 vol='C0*V(plus,c)'
*end of Q computation
*y computation
Gy 0 y cur='v(yd)'
Cy y 0 1
.IC v(y)='y0'
Ry y 0 100meg
*end of y computation
*yd computation
Gyd 0 yd cur='-(4*pi**2*v(y)*((V(y)/y0)**2-1)+gamma*v(yd)+(b*V(plus,minus)/(1+v(y)))**2)'
Cyd yd 0 1
.IC v(yd)='yd0'
Ryd yd 0 100meg
*end of yd computation
.ends BEMS

.option post runlvl=6
Vin 1 0 sin(0,2.8,0.658)
Rin 1 2 1
XMC 2 0 BEMS
.tran 4m 20 16 4m
.probe v(x*.*) i(x*.*)
.end
\end{lstlisting}

\section{SPICE code for model C.4} \label{app_C4}

\noindent{\bf PSpice and LTspice code}

\begin{lstlisting}
**** Bipolar memcapacitive system with threshold C4 ****
*D. Biolek, M. Di Ventra, Y. V. Pershin*
*Reliable SPICE Simulations of Memristors, Memcapacitors and Meminductors, 2013*
*Code for PSpice and LTspice; tested with Cadence PSpice v. 16.3 and LTspice v. 4*
**********************************************************************
.subckt memC_TH plus minus PARAMS:
+ Clow=1p Chigh=100p Cinit=50p beta=70u Vt=3
*model of memcapacitive port
Ec plus minus value={V(Q)/V(x)}
*end of the model of memcapacitive port
*integrator model
Gx 0 x value={fs(V(plus,minus),b1)*ws(v(x),v(plus,minus),b1,b2)}
Raux x 0 100meg
Cx x 0 1 IC={Cinit}
*end of integrator model
*charge computation
GQ 0 Q value={I(Ec)}
CQ Q 0 1
RQ Q 0 100meg
*end of charge computation
*smoothed functions
.param b1=10m b2=1u
.func stps(x,b)={1/(1+exp(-x/b))}
.func abss(x,b)={x*(stps(x,b)-stps(-x,b))}
.func fs(v,b)={beta*(v-0.5*(abss(v+vt,b)-abss(v-Vt,b)))}
.func ws(x,v,b1,b2)={stps(v,b1)*stps(1-x/Chigh,b2)+stps(-v,b1)*stps(x/Clow-1,b2)}
*end of smoothed functions
.ends memC_TH

.options reltol=1u ; use 0.1u for LTspice
*.options method=gear ;use only for LTspice
Vsin 1 0 sin 0 4 50k
Ri 1 2 1m
Xmem 2 0 memC_TH
.tran 0 100u 20u 0.1u
.probe
.end
\end{lstlisting}

\noindent{\bf HSPICE code}

\begin{lstlisting}
**** Bipolar memcapacitive system with threshold C4 ****
*D. Biolek, M. Di Ventra, Y. V. Pershin*
*Reliable SPICE Simulations of Memristors, Memcapacitors and Meminductors, 2013*
*Code for HSPICE; tested with HSPICE RF Version A-2008.03*
**********************************************************************
.subckt memC_TH plus minus
+ Clow=1p Chigh=100p Cinit=50p beta=70u Vt=3
*model of memcapacitive port
Ec plus minus vol='V(Q)/(V(x))'
*end of the model of memcapacitive port
*integrator model
Gx 0 x cur='fs(V(plus,minus),b1)*ws(v(x),v(plus,minus),b1,b2)'
Rx x 0 100meg
Cx x 0 1
.IC v(x)='Cinit'
*end of integrator model
*charge computation
GQ 0 Q cur='I(Ec)'
CQ Q 0 1
RQ Q 0 100meg
*end of charge computation
*smoothed functions
.param b1=10m b2=10u
.param stps(x,b)='1/(1+exp(-x/b))'
.param abss(x,b)='x*(stps(x,b)-stps(-x,b))'
.param fs(v,b)='beta*(v-0.5*(abss(v+Vt,b)-abss(v-Vt,b)))'
.param ws(x,v,b1,b2)='stps(v,b1)*stps(1-x/Chigh,b2)+stps(-v,b1)*stps(x/Clow-1,b2)'
*end of smoothed functions
.ends memC_TH

.option post runlvl=6 delmax=1n
Vsin 1 0 sin(0,4,50k)
Ri 1 2 1
Xmem 2 0 memC_TH
.tran 0.1u 100u
.probe v(x*.*) i(x*.*)
.end
\end{lstlisting}

\section{SPICE code for model L.1} \label{app_L1}

\noindent{\bf PSpice and LTspice code}

\begin{lstlisting}
**** Ideal meminductor L1 ****
*D. Biolek, M. Di Ventra, Y. V. Pershin*
*Reliable SPICE Simulations of Memristors, Memcapacitors and Meminductors, 2013*
*Code for PSpice and LTspice; tested with Cadence PSpice v. 16.3 and LTspice v. 4*
**********************************************************************
.subckt meminductor plus minus params: Llow=1m Lhigh=10m Lini=2m k=10k
.param a={(Lhigh-Lini)/(Lini-Llow)}
*model of meminductive port
.func L(q)={Llow+(Lhigh-Llow)/(a*exp(-4*k*q)+1)}
Ephi phi 0 value={L(V(Q))*I(EL)}
EL plus minus value={ddt(V(phi))}
*end of the model of meminductive port
*integrator model
GQ 0 Q value={I(EL)}
Cint Q 0 1
Raux Q 0 100meg
*end of integrator model
.ends meminductor

Iin 0 1 sin 0 5m 10
XMC 1 0 meminductor
.tran 0 200m 0 200u skipbp ;for LTspice, decrease step ceiling from 200u to 10u
.probe
.end
\end{lstlisting}

\noindent{\bf HSPICE code}

\begin{lstlisting}
**** Ideal meminductor L1 ****
*D. Biolek, M. Di Ventra, Y. V. Pershin*
*Reliable SPICE Simulations of Memristors, Memcapacitors and Meminductors, 2013*
*Code for HSPICE; tested with HSPICE Version A-2008.03*
**********************************************************************
.subckt meminductor plus minus Llow=1m Lhigh=10m Lini=2m k=10k
.param a='(Lhigh-Lini)/(Lini-Llow)'
*model of meminductive port
.param L(q)='Llow+(Lhigh-Llow)/(a*exp(-4*k*q)+1)'
Ephi phi 0 vol='L(V(Q))*I(LL)'
LL plus minus L='L(V(Q))' LTYPE=1
*end of the model of meminductive port
*integrator model
GQ 0 Q cur='I(LL)'
Cint Q 0 1
Raux Q 0 100meg
*end of integrator model
.ends meminductor

.option post runlvl=0 lvltim=1 method=gear
Iin 0 1 sin(0,5m,10)
XMC 1 0 meminductor
.tran 200u 200m
.probe v(x*.*) i(x*.*)
.end
\end{lstlisting}

\section{SPICE code for model L.2} \label{app_L2}

\noindent{\bf PSpice and LTspice code}

\begin{lstlisting}
**** Effective meminductive system L2 ****
*D. Biolek, M. Di Ventra, Y. V. Pershin*
*Reliable SPICE Simulations of Memristors, Memcapacitors and Meminductors, 2013*
*Code for PSpice and LTspice; tested with Cadence PSpice v. 16.3 and LTspice v. 4*
**********************************************************************
.subckt MLsystem plus minus params: L1=1u L2=1u k=0.8 R=1 C=1u
L1 plus minus {L1}
L2 1 3 {L2}
k L1 L2 {k}
R 1 2 {R}
C 2 3 {C}
Raux 3 0 100meg
.ends MLsystem

I 0 in sin 0 1m 100k
XML in 0 MLsystem
Eflux flux 0 value={sdt(v(in))}
.tran 0 60u 40u 0.5n
.probe
.end
\end{lstlisting}

\noindent{\bf HSPICE code}

\begin{lstlisting}
**** Effective meminductive system L2 ****
*D. Biolek, M. Di Ventra, Y. V. Pershin*
*Reliable SPICE Simulations of Memristors, Memcapacitors and Meminductors, 2013*
*Code for HSPICE; tested with HSPICE Version A-2008.03*
**********************************************************************
.subckt MLsystem plus minus L1=1u L2=1u k=0.8 R=1 C=1u
L1 plus minus ‘L1’
L2 1 3 ‘L2’
k L1 L2 ‘k’
R 1 2 ‘R’
C 2 3 ‘C’
Raux 3 0 100meg
.ends MLsystem

.option post
I 0 in sin(0,1m,100k)
XML in 0 MLsystem
Gflux 0 flux cur='v(in)'
Cint flux 0 1
Rx flux 0 100meg
.tran 0.4n 60u 40u 0.5n
.probe v(x*.*) i(x*.*)
.end
\end{lstlisting}

\section{SPICE code for model L.3} \label{app_L3}

\noindent{\bf PSpice and LTspice code}

\begin{lstlisting}
**** Bipolar meminductive system with threshold L3 ****
*D. Biolek, M. Di Ventra, Y. V. Pershin*
*Reliable SPICE Simulations of Memristors, Memcapacitors and Meminductors, 2013*
*Code for PSpice and LTspice; tested with Cadence PSpice v. 16.3 and LTspice v. 4*
**********************************************************************
.subckt memL_TH plus minus PARAMS:
+ Llow=1u Lhigh=100u Linit=50u beta=10meg It=10u
*model of meminductive port
EL plus minus value={ddt(V(phi))}
*forOrCADPSpice 16, the above line can be replaced by EL plus minus F={V(phi)}
*end of the model of meminductive port
*integrator model
Gx 0 x value={fs(I(EL),b1)*ws(v(x),I(EL),b1,b2)}
Raux x 0 100meg
Cx x 0 1 IC={Linit}
*end of integrator model
*flux computation
Ephi phi 0 value={I(EL)*V(x)}
*end of flux computation
*smoothed functions
.param b1=10n b2=1u
.func stps(x,b)={1/(1+exp(-x/b))}
.func abss(x,b)={x*(stps(x,b)-stps(-x,b))}
.func fs(I,b)={beta*(I-0.5*(abss(I+It,b)-abss(I-It,b)))}
.func ws(x,I,b1,b2)={stps(I,b1)*stps(1-x/Lhigh,b2)+stps(-I,b1)*stps(x/Llow-1,b2)}
*end of smoothed functions
.ends memL_TH

.options reltol=1u
*.options method=gear ;use only for LTspice
Isin 0 1 sin 0 12u 50k
Xmem 1 0 memL_TH
.tran 0 100u 40u 0.1u
.probe
.end
\end{lstlisting}

\noindent{\bf HSPICE code}

\begin{lstlisting}
**** Bipolar meminductive system with threshold L3 ****
*D. Biolek, M. Di Ventra, Y. V. Pershin*
*Reliable SPICE Simulations of Memristors, Memcapacitors and Meminductors, 2013*
*Code for HSPICE; tested with HSPICE Version A-2008.03*
**********************************************************************
.subckt memL_TH plus minus
+ Llow=1u Lhigh=100u Linit=50u beta=10meg It=10u
*model of meminductive port
LL plus minus L='V(x)' LTYPE=1
*end of the model of meminductive port
*integrator model
Gx 0 x cur='fs(I(LL),b1)*ws(v(x),I(LL),b1,b2)'
Raux x 0 100meg
Cx x 0 1
.IC v(x)='Linit'
*end of integrator model
*flux computation
Ephi phi 0 vol='I(LL)*V(x)'
*end of flux computation
*smoothed functions
.param b1=10n b2=1u
.param stps(x,b)='1/(1+exp(-x/b))'
.param abss(x,b)='x*(stps(x,b)-stps(-x,b))'
.param fs(I,b)='beta*(I-0.5*(abss(I+It,b)-abss(I-It,b)))'
.param ws(x,I,b1,b2)='stps(I,b1)*stps(1-x/Lhigh,b2)+stps(-I,b1)*stps(x/Llow-1,b2)'
*end of smoothed functions
.ends memL_TH

.option post runlvl=6 KCLTEST delmax=1n
Isin 0 1 sin(0,12u,50k)
Xmem 1 0 memL_TH
.tran 0.1u 100u
.probe v(x*.*) i(x*.*)
.end
\end{lstlisting}

\end{widetext}

\section*{Acknowledgment}
This work has been partially supported by NSF grants DMR-0802830 and ECCS-1202383, the Center for Magnetic Recording Research at UCSD, the SIX Research Center of Sensor, Information and Communication Systems at BUT, and by the development project K217 at UD.
\bibliographystyle{IEEEtran}
\bibliography{IEEEabrv,maze}

\end{document}